# Quantum Spin-Engineering in On-Surface Molecular Ferrimagnets

Wantong Huang[1], Máté Stark[1], Paul Greule[1], Kwan Ho Au-Yeung[1], Daria Sostina[1,2], José Reina Gálvez[3,4], Christoph Sürgers[1], Wolfgang Wernsdorfer[1,2], Christoph Wolf[3,4], Philip Willke[1,*]

[1] Physikalisches Institut, Karlsruhe Institute of Technology (KIT), Karlsruhe, Germany
[2] Institute for Quantum Materials and Technologies, Karlsruhe, Germany
[3] Center for Quantum Nanoscience, Institute for Basic Science (IBS), Seoul, Republic of Korea.
[4] Ewha Womans University, Seoul, Republic of Korea.

* corresponding author: philip.willke@kit.edu,

**The design and control of atomic-scale spin structures constitute major challenges for spin-based quantum technology platforms, including quantum dots, color centers, and molecular spins. Here, we showcase a strategy for designing the quantum properties of molecular spin qubits by combining tip-assisted on-surface assembly with electron spin resonance scanning tunneling microscopy (ESR-STM): We fabricate magnetic dimer complexes that consist of an iron phthalocyanine (FePc) molecule and an organometallic half-sandwich complex formed by the FePc ligand and an attached iron atom, $Fe(C_6H_6)$. The total complex forms a mixed-spin (1/2,1) quantum ferrimagnet with a well-separated correlated ground state doublet, which we utilize for coherent control. As a result of the correlation, the quantum ferrimagnet shows an improved spin lifetime (>1.5 µs) as it is partially protected against inelastic electron scattering. Lastly, the ferrimagnet units also enable intermolecular coupling, that can be used to realize both ferromagnetic or antiferromagnetic structures. Thus, quantum ferrimagnets provide a versatile platform to improve coherent control in general and to study complex magnetic interactions.**

Protecting individual qubits from interaction with the environment is one of the crucial challenges for quantum information processing. For various quantum architectures a plethora of design strategies were developed that alter the properties of the system in a way that makes it resilient to various interactions. One prominent example constitutes the evolution of superconducting qubits, which transitioned from noisy charge qubits to transmon qubits[1], the latter featuring a more robust energy level landscape. Also for spin-based quantum architectures a variety of designs were employed, including clock transitions[2], singlet-triplet qubits in semiconductor quantum dots[3] or chirality-based quantum states[4]. In particular for molecular spin qubits, even larger interacting systems have been proposed and realized[5-7]. Conventionally, these spin systems were enabled by chemical synthesis. An alternative bottom-up route for creating interacting spin structures is on-surface synthesis, monitored and assisted by scanning tunneling microscopy (STM)[8]. This includes various interacting magnetic spin systems that were shown to form complex spin structures[9-16]. However, probing the intrinsic spin properties remains challenging and has up to now mostly been indirectly interrogated via the interaction with the substrate conduction electrons, i.e. the Kondo effect. A viable solution to this problem is to decouple the molecular spins from the metallic substrate via thin insulators[17-19]. This however makes it challenging for on-surface chemistry methods[8].

A direct way to probe spin properties and to obtain insight into their spin dynamics is to use electron spin resonance in a scanning tunneling microscope (ESR-STM)[20]. This allows for probing various spin systems, for instance transition metal atoms[21], alkali atoms[22], rare earth elements[23,24], as well as molecular spins[17,25,26]. Moreover, it permits to control the spin coherently in the time-domain[27,28]. However, short spin lifetimes constitute a major challenge in most spin systems ($T_1$<300 ns[17,25,28]), which also limits their phase coherence time $T_2 \leq 2T_1$. For $T_1$ times, the main limitation remains the scattering with nearby tunneling electrons emanating from the tip and substrate electron baths[27,29,30]. Thus, one viable strategy is to move to thicker layers of the underlying insulator MgO[29], or to employ other routes such as utilizing atomic force microscopy[31]. An alternative strategy is to make the spin systems intrinsically more robust against sources of noise and relaxation by engineering their magnetic interactions.

In this work, we demonstrate how a tip-assisted assembly of a molecular complex leads to a spin system with improved dynamic spin properties compared to the constituents: The complex consists of an iron phthalocyanine (FePc) molecule and an additional Fe atom. The latter forms together with part of the FePc ligand an organometallic half-sandwich complex. We find that the resulting spin system constitutes a mixed spin(1/2,1) Heisenberg quantum ferrimagnet[32]. The ferrimagnet has an improved spin lifetime by a factor of ~5 compared to conventional on-surface spin ½ systems[27,28] reaching here up to $T_1 = 1.6\ \mu s$. This increase is rationalized by employing spin transport calculations, which show that the inelastic scattering channels limiting $T_1$ are suppressed by the correlations in the quantum ferrimagnet. In addition, we demonstrate that multiple complexes can be efficiently coupled either ferromagnetically or antiferromagnetically depending on their relative alignment.

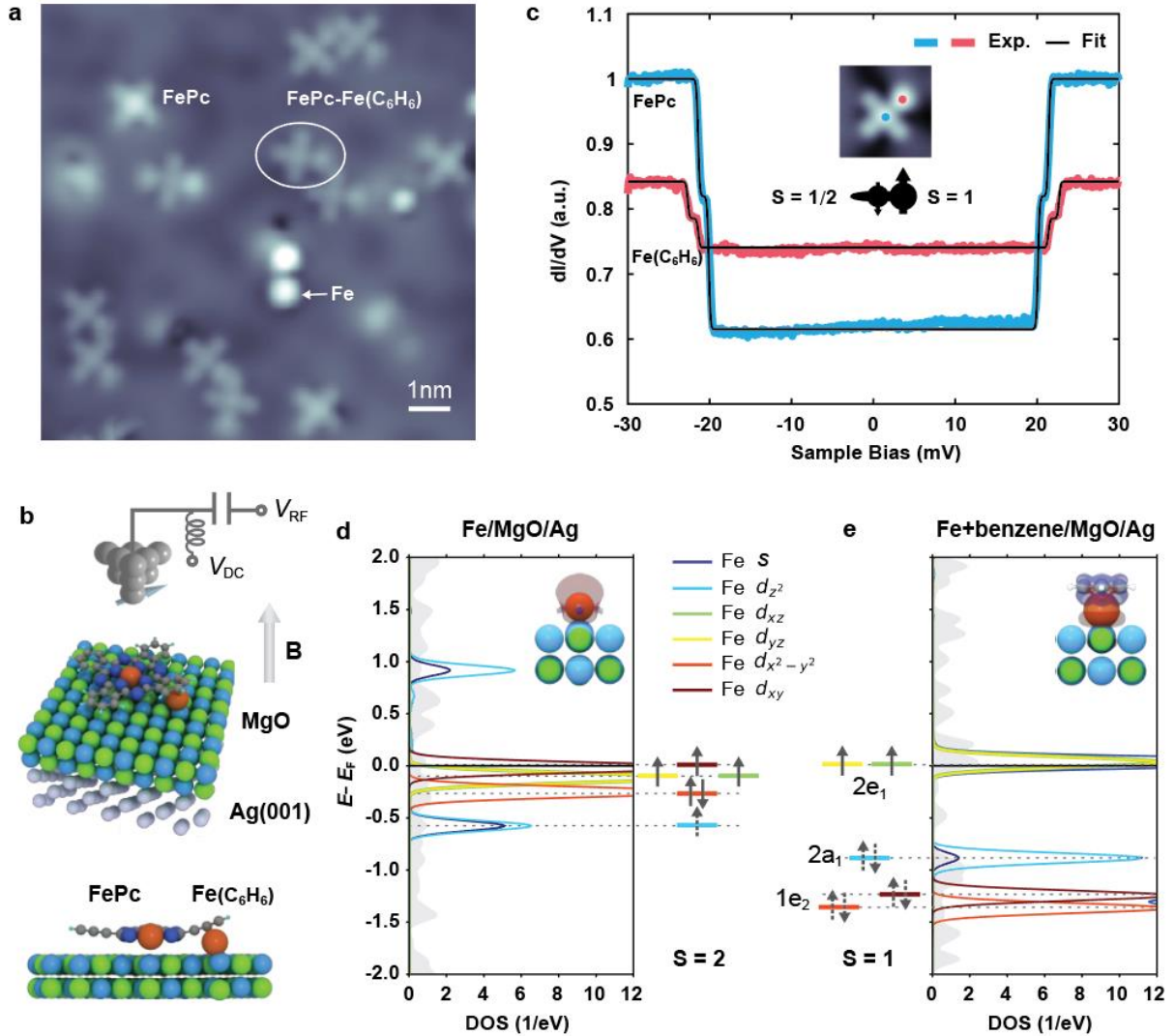

**Figure 1. Engineered spin systems in artificial FePc-Fe(C$_6$H$_6$) complexes. (a)** STM Topographic image showing several molecular FePc-Fe(C$_6$H$_6$) complexes, along with single FePc molecules, Fe atoms and dimers of the complexes (see also Fig. 4) on 2 ML MgO on Ag(001). Additionally, an FePc-[Fe(C$_6$H$_6$)]$_2$ complex is close to the FePc in the upper left corner ($I = 20$ pA, $V_{DC} = -150$ mV, image size: $10.5 \times 10.5$ nm$^2$, $T = 50$ mK). **(b)** Sketch of the ESR-STM experiment: A radio frequency (RF) voltage $V_{RF}$ is applied to the tunnel junction in addition to the DC bias voltage $V_{DC}$. In addition, a magnetic field $B$ is applied out-of-plane. The 3D view and side view image show the stable configuration of the FePc-Fe(C$_6$H$_6$) complex calculated by DFT. Orange spheres show the position of the Fe atoms. **(c)** d$I$/d$V$ spectra acquired at the FePc (blue curve) and Fe(C$_6$H$_6$) site (red curve) as indicated in the inset ($I = 100$ pA, $V_{DC} = 30$ mV, $V_{mod} = 0.25$ mV). The double steps are fitted by inelastic electron spin transport calculations (black lines)[33]. The central sketch illustrates the overall spin structure. **(d)** Projected density of states (PDOS, left) and schematic orbital occupancy (right) for a single Fe atom on 2 ML MgO/Ag(001) obtained from DFT calculation (see inset). It shows the orbitally resolved Fe *s* and 3*d* orbitals and their Lowdin charges, i.e.

projections on atomic states. A strong $4s - 3d_{z^2}$ hybridization (light and dark blue) is found. The dashed arrows indicate shared electrons of partially filled states. **(e)** DOS (right) and orbital occupancy (left) of Fe with an added benzene ring atop (see inset). Here, the Fe states shift in energy in the presence of the ring and are additionally labeled by the respective molecular orbital states. The plot shows projections onto the dominating 3*d* states via Lowdin charges, i.e. atomic orbitals (See Supplementary Section 6). The new ligand framework leaves the 2e₁ orbitals with dominating $d_{xz/yz}$ contribution singly occupied at the Fermi level resulting in S=1.

Figure 1a shows an STM topography of the sample system, that consists of individual Fe atoms and FePc molecules on 2 monolayers (ML) of MgO atop a Ag(001) substrate. FePc molecules were previously shown to be singly charged resulting in a mostly isotropic spin $S = 1/2$ system[25,26], while individual Fe atoms are a spin $S = 2$ system with a large out-of-plane magnetic anisotropy $DS_z^2$ ($D = -4.6$ meV)[34]. From these building blocks we create molecular complexes in which the two Fe atoms are strongly coupled. We employ a simple tip-assisted assembly scheme, in which an FePc molecule is picked up by the STM tip and subsequently dropped atop the Fe atom (Supplementary Fig. S1). In Fig. 1a we show several of these complexes as well as dimers of complexes (discussed in Fig. 4). Lattice site analysis (Supplementary Fig. S2) as well as density functional theory (DFT) calculations (Fig. 1b and Fig. S2b) show that both Fe and FePc adsorb on an oxygen site of MgO with a (2,1) lattice distance. From DFT we also infer that the Fe atom is located underneath the benzene ring ($C_6H_6$) of the attached FePc ligand (Fig. 1b) thus mimicking the structure found in organometallic arene complexes, in particular *half sandwich* or *piano stool* complexes[35]. On surfaces, these kinds of systems have however been only rarely synthesized[36-38]. As a consequence of the shared FePc ligand, the Fe spin under the benzene ring, referred to as Fe($C_6H_6$) in the following, is very close to the FePc central spin. In order to probe the joint magnetic properties of the emerging complexes, we first perform d*I*/d*V* measurements on the Fe($C_6H_6$)- and FePc-site (Fig. 1c). These reveal a double step feature at ~20 meV on both sites. We attribute this to inelastic electron tunneling spectroscopy (IETS) excitations from the magnetic ground state to the excited states. Performing spin transport calculations[33], we can reproduce (black lines in Fig. 1c) both the correct position and intensity of the IETS measurements utilizing a Hamiltonian of the form (see Supplementary Section 3)

$$H = J \cdot \vec{S}_{\text{FePc}} \cdot \vec{S}_{\text{Fe(C6H6)}} + D \cdot S_{z,\text{Fe(C6H6)}}^2 \tag{1}$$

where $J = 14$ meV is the (antiferromagnetic) Heisenberg exchange coupling between the FePc and Fe($C_6H_6$) spins and $D = 1.9$ meV is the out-of-plane magnetic anisotropy of Fe($C_6H_6$). Interestingly, we find that for Fe($C_6H_6$), a spin of $S_{\text{Fe(C6H6)}} = 1$ is required: Using $S_{\text{Fe(C6H6)}} = 2$ as in the case of the isolated Fe atom, we are not able to reproduce the experimental data in Fig. 1c (Supplementary Section 3). This reduction of the Fe spin state is additionally supported by remote magnetic sensing experiments (Supplementary Section 4) as well as IETS measurements on FePc-[Fe(C$_6$H$_6$)]$_2$ complexes (Supplementary Section 5 and Supplementary Fig. S7).

To understand the spin state of Fe(C$_6$H$_6$) we employ DFT calculations (Details see Supplementary Section 6): In Figure 1d we show the projected density of states (PDOS) of an individual Fe atoms' $s$- and $d$-states as discussed elsewhere[39], leading to four orbitals ($d_{xy}, d_{xz}, d_{yz}$ and $d_{z^2}$) close to half-filling and consequently $S_{Fe} = 2$. To rationalize the spin state of Fe(C$_6$H$_6$), we employ a simple model by placing a benzene ring atop the Fe atom (Fig. 1e), as it is the case in the complex and in general for organometallic (half) sandwich complexes[35]. We find that the order of the available states is greatly changed and best described now by those predicted by molecular orbital theory[35]. The molecular orbital diagram of the complex (Supplementary Fig. S8) nicely illustrates, how in particular the e$_1$ states of benzene have a strong overlap with the Fe $d_{xz}$ and $d_{yz}$ orbitals and thus contribute to the stability of the complex.

The resulting two frontier orbitals 2e$_1$ have strong $d_{xz}/d_{yz}$ character and are half-filled, indicating a spin state of $S_{\text{Fe(C6H6)}} = 1$. We note that this resembles to a certain degree the frontier orbitals and spin state found in Nickelocene[40], which consists of two cyclopentadienyl ligands and a Ni ion. We additionally performed DFT calculations of the full FePc-Fe(C$_6$H$_6$) that show a similar molecular orbital formation (See Supplementary Section 6). These also support that the remarkably strong exchange coupling of $J{\sim}14$ meV, the highest observed for spins on MgO, is mediated via the shared FePc ligand. We find that an FePc can host up to 2 Fe(C$_6$H$_6$) complexes (Supplementary Section 5), for which one FePc-[Fe(C$_6$H$_6$)]$_2$ complex is also shown in Fig. 1a.

Figure 2a illustrates again the energy level diagram of the FePc-Fe(C$_6$H$_6$) complex under the action of $D$, $J$, and external magnetic field $B$ as derived from the spin model in Fig. 1c: For small exchange coupling ($J \sim 0$), the out-of-plane anisotropy $D$ lifts the degeneracies of the Fe(C$_6$H$_6$) states. For increasing exchange coupling ($J \neq 0$), two state manifolds form with $S_{tot} \approx \frac{1}{2}$ and $S_{tot} \approx \frac{3}{2}$ which are split by an energy $\frac{3}{2}J$. These transitions between the ground state doublet and the excited states quartet are the transitions observed in the IETS measurements in Fig. 1c and are indicated in Fig. 2a. Figure 2b shows the energy diagram as a function of magnetic quantum number $\langle m_z \rangle$. Since the excited state quartet is further split by $\frac{2}{3}D$, two distinct steps are observed in Fig. 1c. While IETS measurements are well suited to explore the excited state levels, we additionally perform ESR measurements on the complex (Fig. 2c, d) to reveal low lying excitations. Here, a radio frequency (RF) voltage is applied (Fig. 1b) to drive the transition between the two states |0⟩ and |1⟩. Measuring the resonance frequency $f_0$ as a function of external magnetic field $B$ (out of plane) permits to determine the magnetic moment $\mu$ of the complex via the resonance condition $hf_0 = 2\mu B$, where $h$ is Planck's constant. We can perform these measurements on the FePc site (Fig. 2c) and the Fe($C_6H_6$) site (Fig. 2d) and intriguingly, obtain in both cases $\mu \approx 1\ \mu_B$, where $\mu_B$ is the Bohr magneton (Supplementary Section 7 and Supplementary Fig. S12). This suggests that the system acts as one magnetic unit, i.e. a strongly coupled spin system with a ground state that involves both spins: Utilizing again the Hamiltonian in Eq. 1, we plot in Fig. 2e the contribution to the ground state wave function of the coupled system: While in the case of $D \gg J$ the ground state is given by $\left|m_z^{\text{FePc}}; m_z^{\text{Fe(C6H6)}}\right\rangle = \left|\frac{1}{2}; 0\right\rangle$, for increasing exchange interaction a contribution of $\left|-\frac{1}{2}; +1\right\rangle$ is additionally mixed in. In the limiting case of $J \gg D$ the ground state doublet is given as[32]:

$$|0\rangle = \frac{1}{\sqrt{3}}\left|+\frac{1}{2}; 0\right\rangle - \frac{\sqrt{2}}{\sqrt{3}}\left|-\frac{1}{2}; +1\right\rangle$$
$$|1\rangle = \frac{\sqrt{2}}{\sqrt{3}}\left|+\frac{1}{2}; -1\right\rangle - \frac{1}{\sqrt{3}}\left|-\frac{1}{2}; 0\right\rangle$$
(2)

This leads to some noteworthy consequences: First, besides their more complicated form, the two states form a two-level system with $m_z \approx \pm 1/2$ (See Fig. 2b) which are energetically well separated from the higher energy states. This can be roughly understood from the imbalance of the two spins [e.g. (1/2+0) or (-1/2+1) for the |0⟩ state], as it is similarly encountered in ferrimagnetism. Second, the two spin states in Eq. (2) are correlated, since they are not separable anymore, as observed similarly for atomic spin systems[41,42]. To quantify this, we use the *negativity* $\eta$, which is employed as a measure for the pairwise entanglement[41,43] and which becomes non-zero for non-separable states. It approaches $\eta \approx 0.3$ for $J \gg D$ (Fig. 2f)[32]. Thus, the correlated non-separable ground state with $m_z \approx \pm 1/2$ explains well why the same $\mu \approx 1\ \mu_B$ is obtained on both sites in the complex. This is additionally visualized in the spin contrast map in Fig. 2g: Here, the spin density is delocalized over both Fe($C_6H_6$) and FePc sites in the complex.

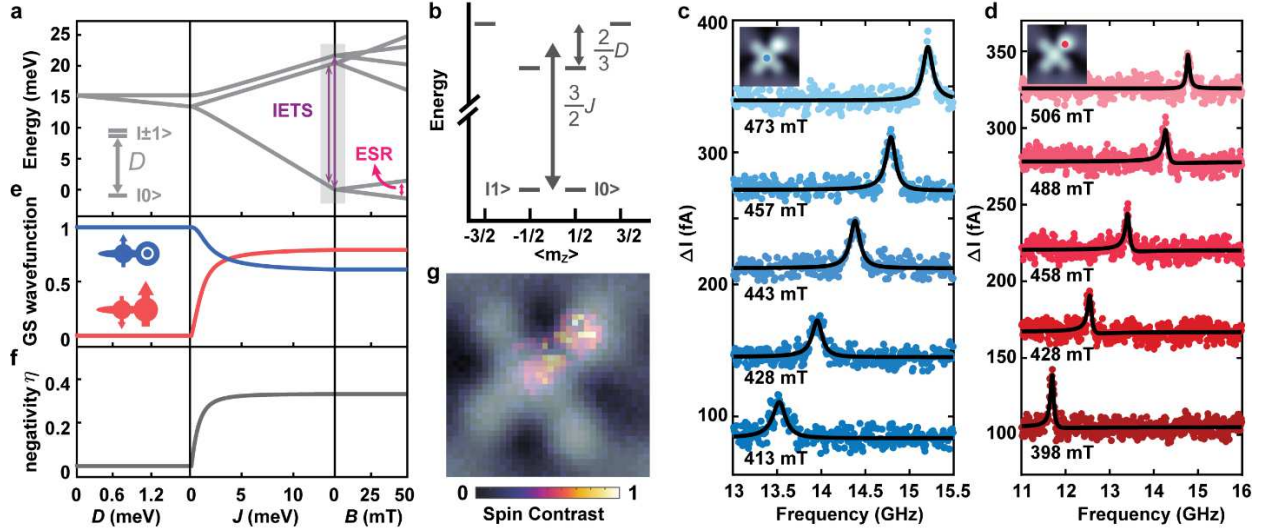

**Figure 2. Electron spin resonance (ESR) of a single FePc-Fe(C$_6$H$_6$) complex in a STM. (a)** Energy level diagram resulting from the simulation of a mixed spin-(1,1/2) Heisenberg ferrimagnet in Eq. 1. The magnetic anisotropy $D$ lifts the degeneracies of a spin 1 at zero field. The exchange coupling ($J$ term) with a spin ½ leads to new $S_{\text{tot}} = 1/2$ and $S_{\text{tot}} = 3/2$ multiplets, which are both further separated by the Zeeman energy in a magnetic field $B$. **(b)** Energy level diagram at zero field shows the states as a function of the $m_z$ expectation value revealing a well-separated ground state doublet. **(c)** ESR-STM measurements on the FePc site ($V = 70$ mV, $I = 50$ pA, $V_{\text{rf}} = 12$ mV) and **(d)** on the Fe(C$_6$H$_6$) site ($V = 25$ mV, $I = 5$ pA, $V_{\text{rf}} = 3$ mV) at different magnetic fields. Linear fits to the slope yield a magnetic moment of $(1.004 \pm 0.012)\,\mu_{\text{B}}$ [FePc site] and $(1.008 \pm 0.007)\,\mu_{\text{B}}$ [Fe(C$_6$H$_6$) site]. The contributions to the ground state wave function **(e)** and negativity $\eta = \sum_j (|\lambda_j| - \lambda_j)/2$ ($\lambda_j$ being an Eigenvalue) **(f)** as a function of $D$, $J$ and $B$. The cartoon insets in (e) illustrate the spin of FePc (left) and Fe(C$_6$H$_6$) (right). **(g)** A spin density map (pseudocolor) with corresponding topography overlaid (gray) of the complex reveals spin contrast on both the FePc and the Fe(C$_6$H$_6$) site. The spin signal was mapped by acquiring a spectroscopic map close to zero bias [|d$^2$I/dV$^2$($V = 2$ mV)|] with a spin-polarized tip and normalized to the highest value (1.7 nm × 1.7 nm, setpoints: $V = 25$ mV, $I = 50$ pA, $V_{\text{mod}} = 2$ mV).

In order to investigate how the unique spin structure of the ferrimagnet complex affects its coherent spin dynamics, we performed pulsed ESR experiments[27]. Figure 3a illustrates a Rabi oscillation measurement (see Supplementary Section 8) on the Fe(C$_6$H$_6$) site [Phase coherence time $T_2^{\text{Rabi}} = (48 \pm 9)$ ns]. Figure 3b shows the power-dependent Rabi oscillations where we observe a linear increase in the Rabi rate $\Omega \propto V_{\text{RF}}$ (Supplementary Fig. S13) as expected for an electric field driven ESR mechanism[25,27]. In the Chevron pattern of the Rabi oscillation (Fig. 3c), the intensity decreases and the oscillation frequency increases as the system is detuned from the resonance frequency. Ramsey fringe measurements (Fig. 3d) also give a dephasing time (~19 ns) of the same order as $T_2^{\text{Rabi}}$ and comparable to those found for single on-surface atomic spins[27]. Thus, they are likely limited by the same sources of decoherence, i.e. $V_{\text{RF}}$-induced tunneling current and fluctuations in the tip magnetic field[27]. However, we here notably achieve a Rabi frequency of $(95.5 \pm 3.2)$ MHz at $V_{\text{RF}} = 60$ mV, which corresponds to a $\pi$-time of $T_\pi = \frac{\pi}{\Omega} = (5.24 \pm 0.17)$ ns. This exceeds the value observed in other spin-1/2 systems, such as Ti[27] and pristine FePc[25], by a factor of ~2 at half the $V_{\text{RF}}$ voltage. We believe that this can originate from multiple sources[44-46], including a larger displacement of the surface spin, a different coupling to the magnetic tip, as well as the change in energetic positions of the molecular orbitals (Fig. 1e, Supplementary Section 8).

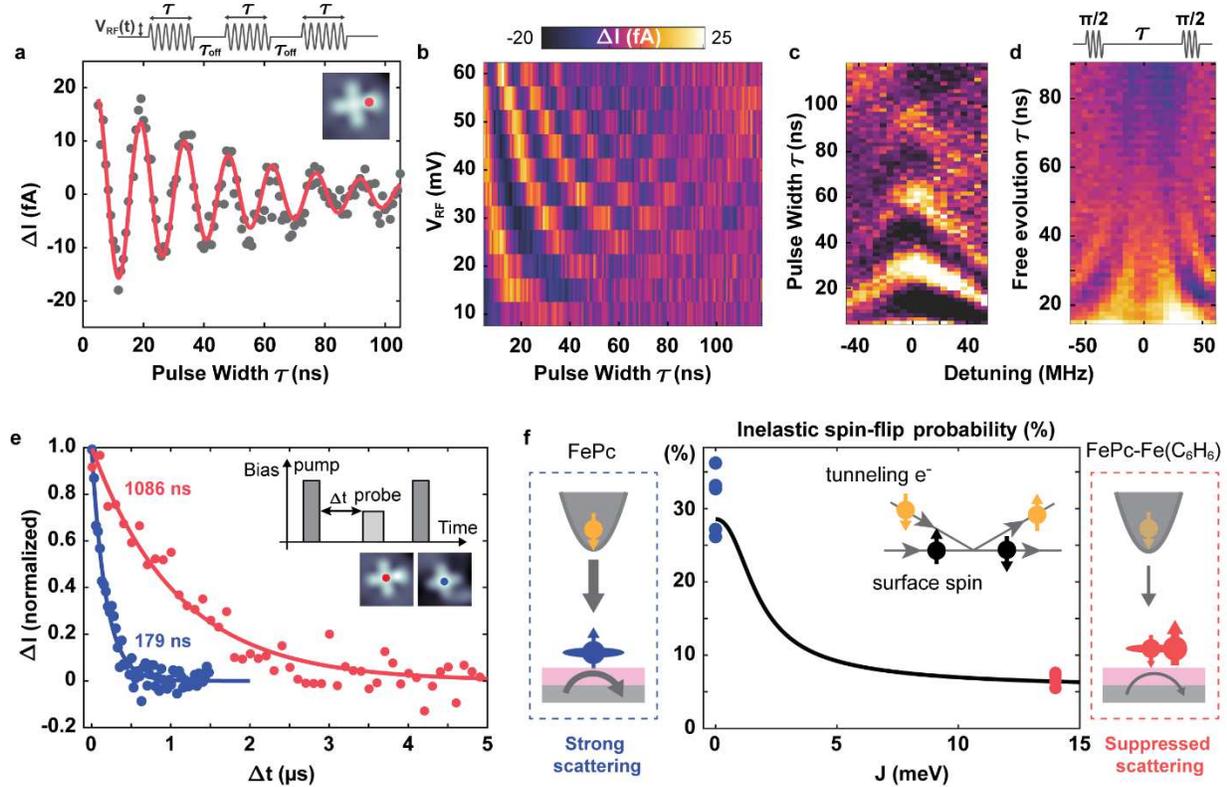

**Figure 3. Spin dynamics and coherent control of the complex. (a)** Rabi oscillation measurement performed at the Fe(C6H6) site in the complex with feedback loop on ($V_{\text{DC}} = -50$ mV, $V_{\text{RF}} = 45$ mV, $I = 4$ pA, $f_0 = 14.512$ GHz, $B = 444$ mT, $\tau_{\text{off}} = 700$ ns). Top: RF pulse scheme of width τ and amplitude $V_{\text{RF}}$. A constant DC voltage $V_{\text{DC}}$ was applied during the sequence as an additional spin initialization and readout following the scheme in Ref.[27]. The solid red curve is a fit to an exponentially decaying sinusoid: $\Delta I = I_0 \sin(\omega\tau + \alpha) \exp\left(-\frac{\tau}{T_2^{\text{Rabi}}}\right)$ for which we obtain $T_2^{\text{Rabi}} \sim 48 \pm 9$ ns. **(b)** Rabi oscillation measurements at different $V_{\text{RF}}$. The linear dependence of the Rabi rate on $V_{\text{RF}}$ is shown in Supplementary Fig. S13. $T_2^{\text{Rabi}} \sim 69 \pm 15$ ns (at $V_{\text{RF}} = 30$ mV). **(c)** Rabi chevron patterns and **(d)** Ramsey fringes with varying

detuning $\Delta = (f - f_0)$. [Setpoint in (c) and (d): $V_{DC} = -60$ mV, $V_{RF} = 60$ mV, $I = 4$ pA, $f_0 = 14.04$ GHz, $B = 473$ mT, $\tau_{cycle} = 450$ ns in (c) and $\tau_{cycle} = 350$ ns in (d). Color scale: [$-10$ fA $\leq \Delta I \leq 15$ fA in (c) and $-7$ fA $\leq \Delta I \leq 6$ fA in (d)]. **(e)** Pump–probe measurements taken on single FePc (blue) and FePc in the complex (red). Solid lines are exponential fits of the form $\Delta I = I_0 \exp\left(-\frac{\tau}{T_1}\right)$, resulting in $T_1 = (1086 \pm 188)$ ns (FePc in complex) and $T_1 = (179 \pm 18)$ ns (single FePc) [parameters: $I_{set} = 25$ pA, $V_{set} = 200$ mV, $B = 600$ mT, $V_{pump} = 80$ mV, $V_{probe} = 40$ mV, $\tau_{pump} = 50$ ns, $\tau_{probe} = 120$ ns, $\tau_{cycle} = 7.8$ μs (FePc in complex), $\tau_{cycle} = 1.7$ μs (single FePc)]. **(f)** Inelastic spin-flip probability as a function of $J$ (black line) derived from spin transport calculations[33] and the states in Eq. (2). The experimental inelastic spin-flip probability extracted from d$I$/d$V$ measurements (see Supplementary Section 10) on 5 pristine FePc molecules [blue dots, (31.2±1.9)%] and from 5 complexes [red dots, (6.7±0.3)%] are shown as well. Inset and sketches to the sides illustrate the spin-flip process in which a tunneling electron exchanges momentum with the on-surface spin.

Additionally, we find a strong enhancement of the relaxation time $T_1$ for which we employ an all electrical pump probe sequence (Figure 3e)[29,47]: Here, a voltage pulse ($V_{Pump}$) pumps the spin system into an excited state while after a varying delay time Δt a second probe pulse ($V_{Probe}$) probes, if the spin still is in the excited state. For direct comparison between FePc-Fe($C_6H_6$) and the pristine FePc, we measure on the FePc site for both cases with the same experimental parameters. From exponential-decay fits we obtain $T_1 = 1086$ ns (ferrimagnet complex), and $T_1 = 179$ ns (pristine FePc). By performing setpoint-dependent $T_1$ measurements, we conclude that the lifetime in both cases is limited by inelastic scattering with tunneling electrons from both the metallic substrate and tip[29] [Supplementary Section 9]. Minimizing the latter, we obtain 1.6 μs (ferrimagnet complex) and 0.4 μs (pristine FePc) in the limit of large tip-surface distance (Supplementary Fig. S14). To explain the ~4-5 times longer lifetime obtained for the complex, we return to its spin model introduced in Fig. 2: The probability for inelastically scattered tunneling electrons results most crucially from the transfer matrix element $|M_{if}^e|^2$, where $M_{if}^e = \langle \varphi_f, \psi_f | \frac{1}{2} \mathbf{S} \cdot \boldsymbol{\sigma} + u | \varphi_i, \psi_i \rangle = m_{if} + u\delta_{if}$[41]. This matrix element connects the initial states in the electron baths $|\varphi_i\rangle$ and spin system $|\psi_i\rangle$ to their final states $|\varphi_f\rangle$ and $|\psi_f\rangle$. $\mathbf{S}$ ($\boldsymbol{\sigma}$) is the spin vector operator of the local spin (tunneling electron). Thus, $m_{if}$ contains the inelastic scattering between the two spins, while $u$ describes spin-conserving Coulomb potential scattering. Thus, the probability for inelastic spin-flip scattering depends on the states $|\psi_{i,f}\rangle$ of the spin system. Figure 3f shows the evolution of this probability as a function of the exchange coupling $J$ when using the states in Eq. 2 for $|\psi_{i,f}\rangle$ (Supplementary Section 10): In the limit of small $J$, the system can be assumed to be in the state of pristine FePc for which we obtain ~29%. This drops to ~6% for the complex as the spin system is transitioning into the correlated ground state. This suggests a decrease by ~4-5, which matches well with the results for the pump-probe data. In addition to the simulation, we can evaluate the suppressed inelastic scattering from spin-polarized IETS d$I$/d$V$ measurements (Supplementary Section S10). The probabilities of inelastic scattering obtained from these measurements are additionally shown in Fig. 3f and agree well with the spin transport calculations.

The results of Fig. 3 demonstrate that the FePc-Fe($C_6H_6$) complex allows for coherent manipulation with an enhanced spin lifetime as well as a fast Rabi rate. We believe that the latter also benefits from the suppressed inelastic scattering, since this allows us to perform measurements at closer tip-sample distances and thus stronger tip fields without significantly reducing relaxation and coherence. In general, we think that in particular the protection makes quantum ferrimagnets a promising system for quantum sensing, simulation, or information processing: The exchange-like interaction between the complex and the tunneling electrons is similar to sources of relaxation in other quantum architectures such as flip-flop processes with nuclear spins.

However, a crucial ingredient for creating more complex spin structures is the creation of interacting spin systems from individual building blocks. Figure 4a shows a dimer consisting of two closeby FePc-Fe($C_6H_6$) complexes (several more are shown in Fig. 1). A direct advantage of the complex for spin-spin-coupling is the position of the Fe($C_6H_6$) spin at the FePc ligand. This arrangement facilitates coupling with the adjacent complex due to a shorter distance between spins and leads to much larger J coupling than for dimers of pristine FePc[26]. The coupling scheme is illustrated in Fig. 4b, with the intramolecular coupling $J_1 = 14$ meV as used before and the weaker intermolecular coupling $J_2^{eff}$ (between ferrimagnets). Treating for now the two ferrimagnet complexes as effective spin ½ systems leads to a Hamiltonian of the form[26]

$$H = H_1 + H_2 + D_{dip} \cdot (\boldsymbol{S}_1 \cdot \boldsymbol{S}_2 - 3[\boldsymbol{S}_1 \cdot \hat{\boldsymbol{r}}] \cdot [\boldsymbol{S}_2 \cdot \hat{\boldsymbol{r}}]) + J_2^{eff} \cdot \boldsymbol{S}_1 \cdot \boldsymbol{S}_2 \tag{3}$$

Where $H_1 = g_1 \mu_B \boldsymbol{S}_1 \cdot (\boldsymbol{B} + \boldsymbol{B}_{\text{tip}})$ marks the single-site Zeeman energy of the left and $H_2 = g_2 \mu_B \boldsymbol{S}_2 \cdot \boldsymbol{B}$ the term of the right complex. Here, $D_{dip}$ is the effective magnetic dipolar coupling between two spins.

To investigate the interacting spin system, we perform ESR measurements on one of the spins for varying tip magnetic field $B_{\text{tip}}$. Since $B_{\text{tip}}$ contributes in a Zeeman-like way, but only to one of the complexes[26], it allows to tune the coupled spin system through an avoided level crossing as shown in Fig. 4c: Here, we depict the energy level diagram of two coupled spin-½ as a function of $B$ and $B_{tip}$. We additionally highlight the dominating ESR transitions below ($f_3$ and $f_4$) and above ($f_1$ and $f_2$) the point of no detuning $\delta = g_1 \mu_B (B + B_{tip}) - g_2 \mu_B B = 0$.

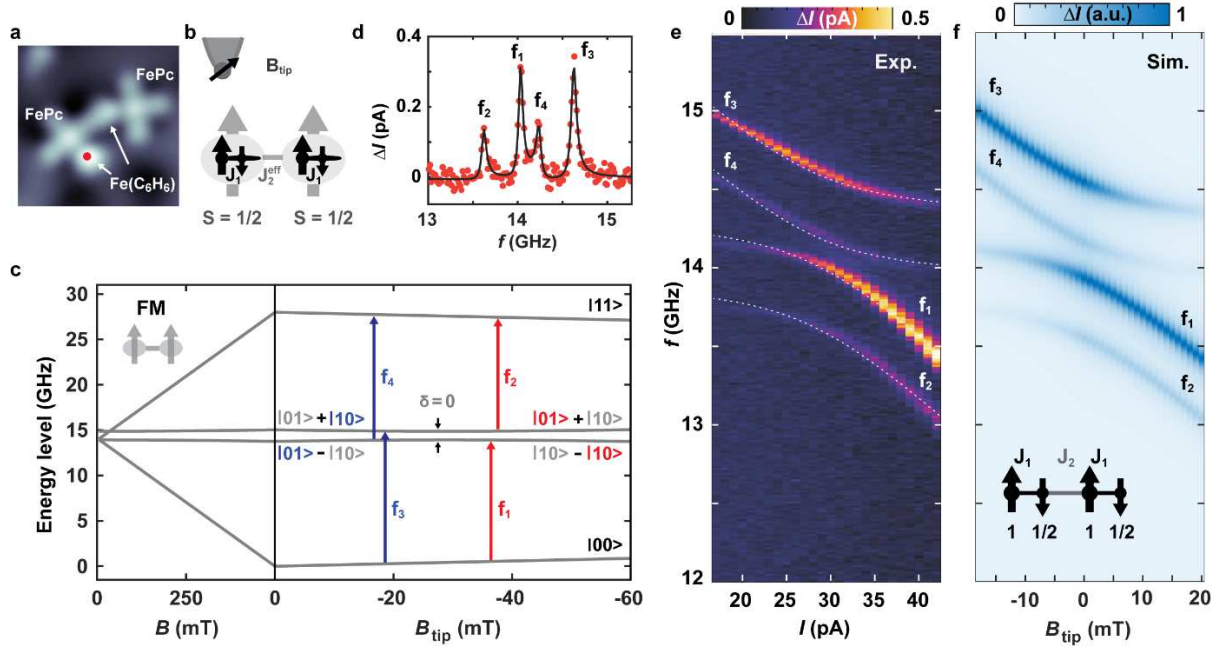

**Figure 4**. **Spin coupling in a dimer of ferrimagnet complexes (a)** Topographic image of two coupled complexes ($I = 10$ pA, $V_{DC} = 100$ mV, image size: $3 \times 3$ nm²) with a spacing of 1.04 nm [(3,2) MgO lattice sites]. **(b)** Illustration of the magnetic coupling within the spin structure: The small (large) black arrows represent the FePc [Fe(C$_6$H$_6$)] spins. Within each individual complex two spins are antiferromagnetically coupled with strength $J_1$. Since also the FePc and Fe(C$_6$H$_6$) in different complexes couple antiferromagnetically, the overall coupling between the complexes is ferromagnetic with coupling strength $J_2^{eff}$. **(c)** Energy level diagram of two coupled spin ½ in the presence of $B$ and $B_{tip}$. The detuning $\delta$ quantifies the distance from the avoided level crossing. **(d)** ESR frequency sweep measurement at a fixed $B_{tip}$ showing four peaks $f_1$ to $f_4$, corresponding to the transitions labelled in (c). All data are measured on the Fe(C$_6$H$_6$) site position marked as a red dot in (a). ($V_{DC} = 60$ mV, $I_{set} = 29$ pA, $V_{RF} = 12$ mV, $B = 473$ mT). Black line is a fit of four Lorentzians to the data. **(e)** ESR frequency sweep measurements at different setpoint currents ($B_{tip}$) showing an avoided level crossing ($V_{DC} = 60$ mV, $V_{RF} = 12$ mV, $B = 473$ mT). The dotted white curves represent fitted ESR transitions using a 2 spin model, with $J_2^{eff} = -531$ MHz and $D_{dip} = 131$ MHz, $g_1 = 2$, $g_2 = 2.01$. **(f)** Simulation of the coupled spin system using a four spin model (inset) of spins (1, 1/2, 1, 1/2) with antiferromagnetic coupling strengths $J_1$ and $J_2$. The fitting parameters are: $J_1 = 14.65$ meV, $J_2 = 1.2$ GHz, $g_{FePc} = 2$, $g_{Fe} = 2$.

At the latter, the single site Zeeman energies for the complex on the left ($H_1$) and right ($H_2$) are equal. These kinds of measurements have been realized before on regular coupled spin ½ systems, i.e. Ti atoms[48,49] and pristine FePc[26]. We show in the following that also the more complex ferrimagnetic spins can be coupled as well and in fact constitute more flexible magnetic building blocks. An ESR spectrum measured on the Fe($C_6H_6$) site within the left complex reveals four ESR peaks (Fig. 4d), corresponding to those labeled in Fig. 4c. Figure 4e shows an ESR map measurement as a function of frequency $f$ and tip setpoint current $I$. The latter is in good approximation $\propto B_{\text{tip}}$[46] and can tune the frequency and intensity of the four ESR transitions. The evolution shows good agreement with the predictions of an effective two-spin model (Fig. 4e, see Supplementary Section 11) which suggests that the two complexes can be described as two coupled effective spin ½ with intermolecular coupling strength $J_2^{\text{eff}}$. This makes them suitable building blocks for larger spin systems. However, a notable difference compared to previous works on atomic and molecular spins is that $J_2^{\text{eff}}$ is negative, constituting a ferromagnetic (FM) exchange interaction between the two effective spins. This can be rationalized when considering the internal structure of each complex (Fig. 4b): Here, the coupling between all adjacent spins remains antiferromagnetic, as it is most frequently found for superexchange interaction. However, the two ferrimagnet complexes can also be coupled antiferromagnetically (AFM, $J_2^{\text{eff}} > 0$) when positioning the two Fe($C_6H_6$) centers in each complex closest together (Supplementary Figs. S18-19 and Supplementary Section 11).

To capture minute details in the position and intensity of the peaks (see Supplementary Section 11), we employ a simulation of the full four-spin system (Fig. 4f), where all magnetic dipolar couplings are calculated based on the distances between all spin centers (Supplementary Fig. S20).

Our measurements highlight how both atomic and molecular building blocks can be combined to tailor new molecular spin structures. A crucial ingredient is the bond formation between the Fe and the benzene ring, which instead of using tip-assisted on-surface assembly could also be self-assembled by other on-surface chemistry techniques[50], or directly synthesized in fully chemically derived molecular ferrimagnets.

In particular the intricate connection between the emerging correlated spin state of the complex and the improvement in its dynamic properties reveals a key strategy to protect spins on surfaces from the main source of relaxation, i.e. scattering by nearby electron spins. We believe that combining this protection with an increased MgO layer thickness[29] and a remote spin readout[28] will allow to further enhance $T_1$ and $T_2$. For a remote spin readout, the demonstration of efficient spin-spin coupling (Fig. 4) is an important prerequisite, which also offers the opportunity to design larger spin structures or quantum simulators with both AFM and FM coupling. Lastly, we note that the concept of strongly coupled quantum ferrimagnets employed here is not unique to assembled molecular spins on surfaces: The ingredients for the spin system are quite universal and could also be realized by spins in other quantum architectures including quantum dots, color centers or fully chemically derived molecular ferrimagnets.

**Materials and Methods**

**Sample preparation**

All sample preparation was carried out in-situ at a base pressure of < 5 × $10^{-10}$ mbar. The Ag(001) surface was prepared through several cycles of argon-ion sputtering and annealing through e-beam heating. For MgO growth, the sample was heated up to 510°C and exposed to an Mg flux for 10 minutes in an oxygen environment at $10^{-6}$ mbar leading to an MgO coverage of ~50% and layer thicknesses ranging from 2 to 5 monolayers. FePc was evaporated onto the sample held at room temperature using a home-built Knudsen cell at a pressure of 9 × $10^{-10}$ mbar for 90 seconds. Electron-beam evaporation of Fe was carried out for 21 seconds onto the cold sample. We determined the thickness of MgO layers through point-contact measurements on single Fe atoms [29]. All experiments were carried out using a Unisoku USM1600 STM inside a homebuilt dilution refrigerator with a base temperature of 50 mK. An effective spin temperature of ~300 mK was estimated from ESR measurements of Fe dimers. Here, the intensities of ESR peaks of the coupled electron spin states depend on temperature[51], which we take as an estimate of the Boltzmann distribution in the experiment.

**Pulsed electron spin resonance measurements**

Spin-polarized tips were prepared by picking up individual Fe atoms from the surface. The spin polarization was tested through the asymmetry in d$I$/d$V$ spectra of FePc with respect to voltage polarity. Magnetic tips showing a high spin contrast were subsequently tested in continuous-wave (CW) ESR-STM measurements. The RF voltage was applied on the tip-side of the junction using an RF generator (Rohde & Schwarz SMB100B). The RF voltage was combined with the DC tunnel bias using a Diplexer (Marki Microwave MDPX-0305). Note that while the bias voltage was applied to the STM tip, all bias signs were inverted in the manuscript to follow the conventional definition of bias voltage with respect to the sample bias.

For pulsed ESR measurements, we followed the manipulation scheme introduced in Ref.[27]. An arbitrary waveform generator (Zurich Instruments HDAWG) gated the output of the RF generator to generate the desired pulsed ESR scheme. We used a digital lock-in amplifier (Stanford

Research Systems SR860) to read out the ESR signal using an on/off modulation scheme at 323 Hz. During the Rabi or Ramsey measurements, the feedback loop was set to low gain values. In total, these measurements took ~3 seconds per data point, so that a typical Rabi or Ramsey measurement including 3 iterations of 100 points took around 15 minutes. All measurements were conducted by using schemes described in Ref. [27], in which a DC background current is applied in lock-in A and B cycles. While RF pulses were only applied in lock-in A cycles. Linear backgrounds that originate from an additional rectification for increasing pulse lengths[27] have been subtracted.

For pump probe measurements in Fig. 3 and Supplementary Fig. S14, we altered the tunnel current by changing the bias voltage to zero once the feedback loop had been disengaged. Consequently, all measurements have been performed at constant tunnel conductance and constant tip-atom distance, which in particular helped to keep the influence of the tip field constant. For each measurement, the lockin A cycles include both pump and probe pulse voltages while the lock-in B cycles are left empty, with no DC voltages applied.

**Associated Content**

Supporting Information: Building FePc-[Fe($C_6H_6$)]$_x$ complexes and chains, Lattice site analysis, Spin Hamiltonian and IETS d*I*/d*V* spectra simulations, Magnetic sensing, d*I*/d*V* simulation of an FePc-[Fe($C_6H_6$)]$_2$ complex, Details of DFT calculations, ESR and fitting, Rabi rate as a function of $V_{RF}$, Conductance-dependent spin lifetime, Spin transport simulations and enhanced spin lifetime, ESR transitions in Heisenberg two-spin system with the tip field detuning effect

**Acknowledgments**

P.W. acknowledges funding from the Emmy Noether Programme of the DFG (WI5486/1-1) and financing from the Baden Württemberg Foundation Program on Quantum Technologies (Project AModiQuS). P.G. and P.W. acknowledges financial support from the Hector Fellow Academy (Grant No. 700001123). C.W. and J.R. acknowledged support from the Institute for Basic Science (IBS-R027-D1).

**Author contributions.** P.W. and W.H. conceived the research. W.H., M.S., P.G. K.H.A.Y., D.S., C.S., W.W. and P.W. set up the experiment and conducted the measurements. W.H., M.S., P.G., K.H.A.Y. and P.W. analyzed the experimental data. C.W. performed the DFT calculations. J.R participated in the modeling of the spin systems. W.H. and P.W. wrote the manuscript with input from all authors. W.W. and P.W. supervised the project.

**Notes.**

The authors declare no competing financial interests.

# Supplementary Information

# Quantum Spin-Engineering in On-Surface Molecular Ferrimagnets

Wantong Huang[1], Máté Stark[1], Paul Greule[1], Kwan Ho Au-Yeung[1], Daria Sostina[1,2], José Reina Gálvez[3,4], Christoph Sürgers[1], Wolfgang Wernsdorfer[1,2], Christoph Wolf[3,4], Philip Willke[1,*]

[1] Physikalisches Institut, Karlsruhe Institute of Technology (KIT), Karlsruhe, Germany
[2] Institute for Quantum Materials and Technologies, Karlsruhe, Germany
[3] Center for Quantum Nanoscience, Institute for Basic Science (IBS), Seoul, Republic of Korea.
[4] Ewha Womans University, Seoul, Republic of Korea.

* corresponding author: philip.willke@kit.edu,

## Table of Contents

## 1. Building FePc-[Fe(C$_6$H$_6$)]$_x$ complexes

The FePc-Fe(C$_6$H$_6$) complexes were created using the atom manipulation scheme sketched in Fig. S1a. First, the tip is positioned above one ligand of FePc. The tip is then moved ~600 pm closer to the molecule and a bias voltage of 850 mV is applied. Next, the tip is retracted from the surface with the FePc sticking to its apex. To drop off the molecule, we used two methods: 1) Perform Z-Spectroscopy starting with setpoints of 10pA and $-10$mV, gradually moving the tip $250 - 330$ pm towards the Fe atom. 2) Scan fast across the Fe atom at closer distance ($V_{set} = -40$ mV, $I_{set} = 1$ nA and scan speed of 300 nm/s). Figure S1b and S1c show the topographic images before and after building an FePc-Fe(C$_6$H$_6$) complex, respectively. Fig. S1d,e show additional cross sections across the complex, in which the protrusion at the position of the Fe atom becomes greatly visible. In Fig. 1a in the main text, we also showed multiple dimers of complexes and an FePc-[Fe(C$_6$H$_6$)]$_2$ complex, which consists of two Fe(C$_6$H$_6$) and one FePc (see Supplementary Section 5). These were built using similar atom manipulation schemes. For FePc-[Fe(C$_6$H$_6$)]$_2$ complexes, we first locate two closed Fe atoms and then drop the molecule on one of them, potentially forming an FePc-[Fe(C$_6$H$_6$)]$_2$. For two coupled FePc-Fe(C$_6$H$_6$) complexes, we build two complexes separately first and then move one towards the other. To move the complex, the tip is positioned between two ligands ($V_{set} = -100$ mV, $I_{set} = 20$ pA), next moved $200 - 400$ pm closer to the surface and finally a bias voltage of $500 - 750$ mV was applied.

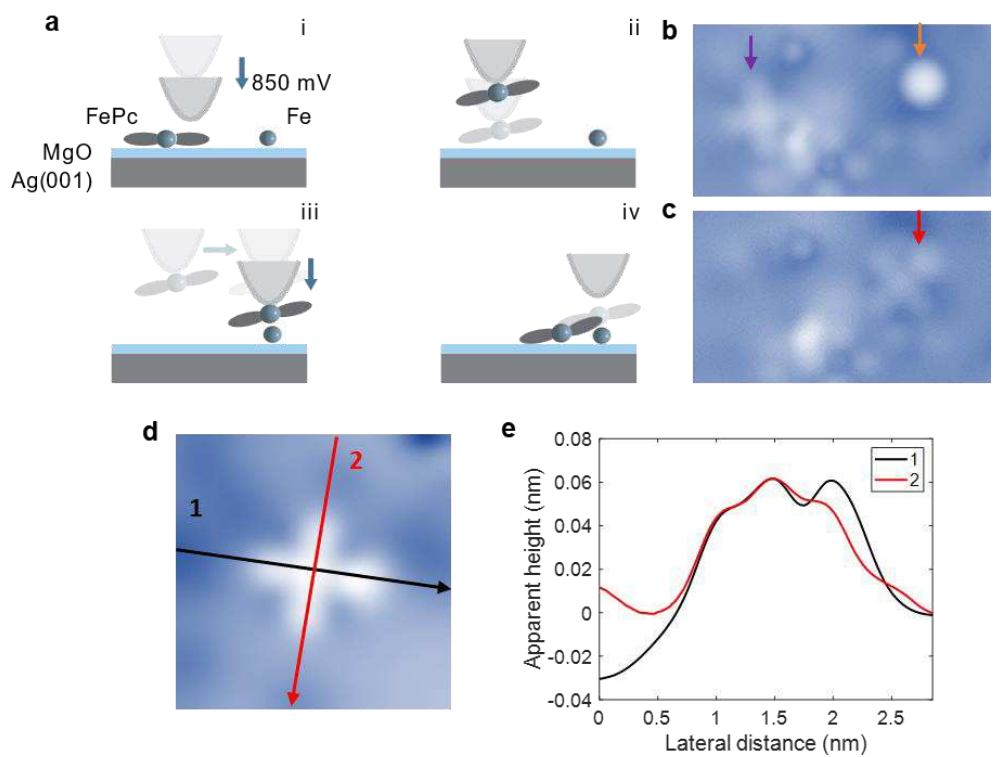

**Supplementary Fig. S1. a,** Schematic illustrating the manipulation procedures involved in building FePc-Fe($C_6H_6$) complexes. **b,** Topographic image consisting of one Fe atom (orange arrow) and one FePc molecule (purple arrow) on 2ML MgO ($V = 100$ mV, $I = 20$ pA). **c,** Topographic image afterwards with a newly formed FePc-Fe($C_6H_6$) complex (red arrow). **d,** STM topography image ($V = 100$ mV, $I = 20$ pA) and **e,** the corresponding line profiles comparing the ligands of the FePc-Fe($C_6H_6$) complex with (black) and without (red) the Fe adatom underneath the benzene ring.

## 2. Lattice site analysis

The FePc-[Fe($C_6H_6$)]$_x$ (x=1 or 2) complexes adsorbed on 2ML MgO/Ag(001) tend to adopt specific site configurations. The precise positioning of the Fe atoms and FePc molecules with respect to the underlying MgO lattice is determined by atomic resolution images of the bare MgO area. In addition, the position of nearby individual Fe atoms, which are adsorbed atop a MgO oxygen lattice site[1], is taken as a reference of the lattice. In Fig. S2a (Fig. S2c), the oxygen lattice is indicated by a white grid overlaid on the FePc-Fe($C_6H_6$) (FePc-[Fe($C_6H_6$)]$_2$) complex structure. The Fe atoms (marked in red) are situated at a distance of 0.64 nm (2×1 oxygen lattice sites) from the center of FePc. Consequently, both Fe and FePc are still adsorbed atop oxygen sites (Fig. S2b), as it is the case for the isolated atom and molecule[1,2]. These specific positions are used as the basis for our DFT calculations.

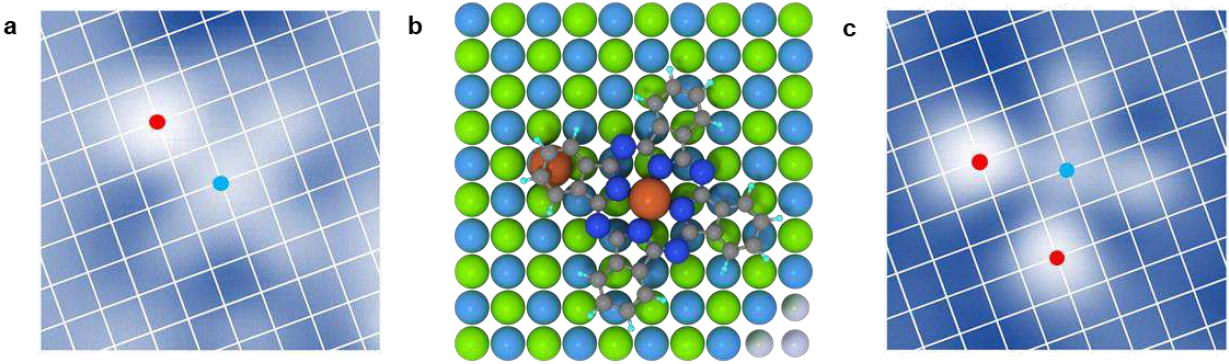

**Supplementary Fig. S2. FePc-[Fe(C$_6$H$_6$)]$_x$ complexes binding site. a,** Topographic image of an FePc-Fe(C$_6$H$_6$) complex with MgO lattice overlaid (2nm × 2 nm, setpoint: $V$ = 100 mV, $I$ = 20 pA). **b,** Top view image shows the stable configuration of the FePc-Fe(C$_6$H$_6$) complex calculated by DFT. **c,** Topographic image of an FePc-[Fe(C$_6$H$_6$)]$_2$ complex with the MgO lattice overlaid (2nm × 2 nm, setpoint: $V$ = -200 mV, $I$ = 20 pA). The positions of Fe atoms and molecule centers are labeled by red and blue dots, respectively. The position of oxygen atoms in the MgO lattice is indicated by a white grid.

### 3.    Spin Hamiltonian and IETS dI/dV spectra simulations

In this section, we introduce the IETS d$I$/d$V$ spectra simulations and investigate in detail the spin Hamiltonian described for the FePc-[Fe(C$_6$H$_6$)]$_2$ complexes in the main text. When applying the external magnetic field $B$ in z-direction, we consider the following Hamiltonian for the complex:

$$H = J \cdot \vec{S}_{\text{FePc}} \cdot \vec{S}_{\text{Fe(C6H6)}} + D \cdot \hat{S}^2_{z,\text{Fe(C6H6)}} - g_1 \mu_B B \hat{S}_{z,\text{Fe(C6H6)}} \\ - g_2 \mu_B B \hat{S}_{z,\text{FePc}} \qquad (3)$$

Where $J$ is the Heisenberg exchange coupling, $D$ is the out-of-plane zero-field splitting term of the Fe(C$_6$H$_6$) spin, $\mu_B$ is the Bohr magneton, $B$ is the external perpendicular magnetic field and $g_1$ and $g_2$ are g-factors of the Fe(C$_6$H$_6$) and FePc spins, respectively. In our case, $S_{\text{FePc}} = 1/2$ and $S_{\text{Fe(C6H6)}} = 1$. The energies and eigenstates of this mixed spin-(1/2,1) system can be expressed analytically[3]. The eigen-energies are:

$$E_{1,2} = \frac{1}{2}[J + 2D \mp (g_1 \mu_B B + 2g_2 \mu_B B)], \qquad (4)$$

$$E_{3,4} = -\frac{1}{4}(J - 2D + 2g_2\mu_B B) \mp \frac{1}{4}\sqrt{[J - 2D - 2(g_1\mu_B B - g_2\mu_B B)]^2 + 8J^2}, \tag{5}$$

$$E_{5,6} = -\frac{1}{4}(J - 2D - 2g_2\mu_B B) \mp \frac{1}{4}\sqrt{[J - 2D + 2(g_1\mu_B B - g_2\mu_B B)]^2 + 8J^2}, \tag{6}$$

whereas the corresponding eigenvectors are

$$\psi_{1,2} = |\pm\frac{1}{2}, \pm 1>, \tag{7}$$

$$\psi_{3,4} = c_1^{\mp}\left|+\frac{1}{2}, 0\right> \mp c_1^{\pm}\left|-\frac{1}{2}, 1\right>, \tag{8}$$

$$\psi_{5,6} = c_2^{\pm}\left|+\frac{1}{2}, -1\right> \mp c_2^{\mp}\left|-\frac{1}{2}, 0\right>. \tag{9}$$

The probability amplitudes in the last four eigenvectors (8) and (9) are:

$$c_1^{\pm} = \frac{1}{\sqrt{2}}\sqrt{1 \pm \frac{J - 2D - 2(g_1\mu_B B - g_2\mu_B B)}{\sqrt{[J - 2D - 2(g_1\mu_B B - g_2\mu_B B)]^2 + 8J^2}}},$$

$$c_2^{\pm} = \frac{1}{\sqrt{2}}\sqrt{1 \pm \frac{J - 2D + 2(g_1\mu_B B - g_2\mu_B B)}{\sqrt{[J - 2D + 2(g_1\mu_B B - g_2\mu_B B)]^2 + 8J^2}}}. \tag{10}$$

For $J \gg D, g_1\mu_B B, g_2\mu_B B$, the factors $c_{1,2}^+ \cong \sqrt{\frac{2}{3}}$, $c_{1,2}^- \cong \sqrt{\frac{1}{3}}$.

At zero field $B = 0$ T (and $J \gg D$), this simplifies the energies to:

$$E_{1,2} = \frac{1}{2}(J + 2D), \tag{11}$$

$$E_{3,5} = -\frac{1}{4}(J - 2D) - \frac{1}{4}\sqrt{(J - 2D)^2 + 8J^2} \cong -J + \frac{2}{3}D, \tag{12}$$

$$E_{4,6} = -\frac{1}{4}(J - 2D) + \frac{1}{4}\sqrt{(J - 2D)^2 + 8J^2} \cong \frac{J}{2} + \frac{1}{3}D, \tag{13}$$

This approximation nicely illustrates that the ground state energy is given by $E_{3,5}$ (assuming antiferromagnetic exchange, i.e. $J > 0$). The wave functions of the ground state doublet are given as $\psi_{3,5}$:

$$|0\rangle = \psi_3 = \frac{1}{\sqrt{3}}\left|+\frac{1}{2};0\right\rangle - \frac{\sqrt{2}}{\sqrt{3}}\left|-\frac{1}{2};+1\right\rangle$$

$$|1\rangle = \psi_5 = \frac{\sqrt{2}}{\sqrt{3}}\left|+\frac{1}{2};-1\right\rangle - \frac{1}{\sqrt{3}}\left|-\frac{1}{2};0\right\rangle$$

(14)

The corresponding energy level diagram is shown in Fig. S3a (see also Fig. 2b). Two state manifolds form with $S_{tot} \approx 1/2$ and $S_{tot} \approx 3/2$ which are separated by an energy $\sim \frac{3}{2} \cdot J$. The excited state quartet is further split by $2/3 \cdot D$.

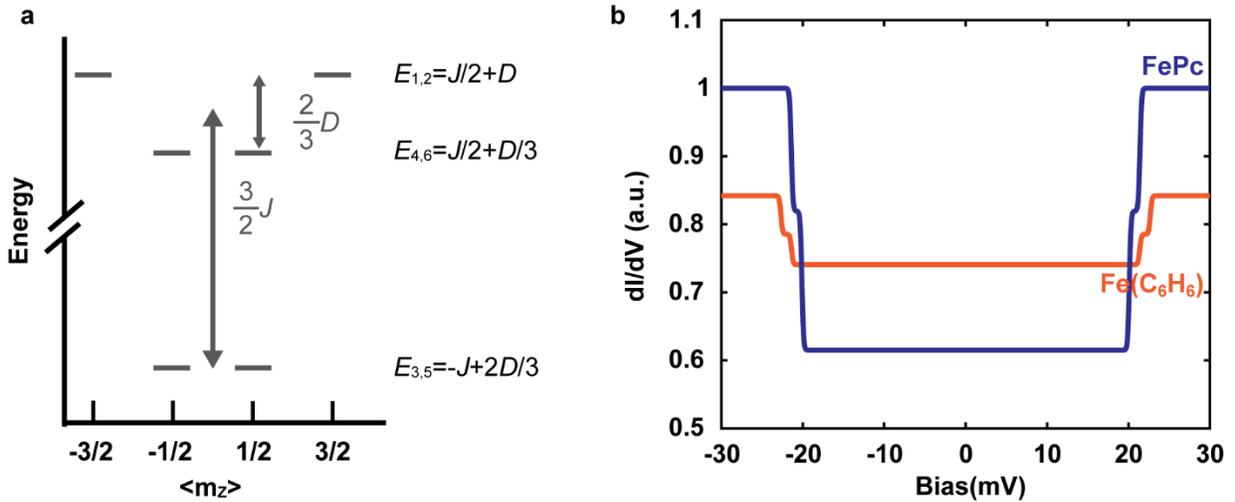

**Supplementary Fig. S3. a,** Energy level diagram of a mixed-spin (1/2,1) Heisenberg quantum ferrimagnet. **b,** Simulated IETS dI/dV spectra with S = ½ for FePc, S = 1 for Fe($C_6H_6$). Fitting parameters: For Fe($C_6H_6$): J = 14.65 meV, D = 1.9 meV, U = 0.945. For FePc: J = 13.8 meV. U = 0.494. T=0.8 K.

In the IETS simulations, we take the solution of the Hamiltonian described above to calculate the resulting tunneling current by using the spin simulation code developed in Ref.[4,5]. The latter considers contributions to the tunneling current stemming from 1) potential scattering $U$ and 2) inelastic scattering of the tunneling electrons. The values of parameters $D, J$ and $U$ are adjusted for Fe($C_6H_6$) and FePc in the complex to match the step heights and positions in the experiment: The simulated IETS dI/dV spectra shown in Fig. S3b (see also Fig. 1c) reproduce the double step feature found in the experimental data well.

The IETS calculations and the resulting d*I*/d*V* spectra are crucially influenced by the assumptions made in the initial spin Hamiltonian. As such, the IETS spectrum plays an important role to determine the spin state of the Fe(C$_6$H$_6$) in the complex. We screened a wide variety of cases and found that in particular the double step feature (along with constraints imposed by the results of the ESR measurements, remote sensing experiments as well as DFT calculations), can only be explained assuming $S_{\text{FePc}} = ½$ and $S_{\text{Fe(C6H6)}} = 1$ as shown in Fig. S3. In the following, we will exemplarily illustrate, how other reasonable spin models are not in agreement with the data.

1) **FePc: S = 1/2, Fe(C$_6$H$_6$): S = 2**

A first guess for the FePc spin state and the Fe spin would be adapting those of the individual spin systems, i.e. $S_{\text{FePc}} = ½$ and $S_{\text{Fe(C6H6)}} = 2$. In this case, FePc and Fe(C$_6$H$_6$) retain their spin states as individual FePc (charged by one electron) and Fe on MgO. For the simulation, we initially assume that Fe has the same magnetic anisotropy ($D = -4.7 \text{ meV}^6$) as for isolated Fe atoms and that there is Heisenberg exchange coupling between the two spins. For ferromagnetic coupling ($J < 0$), there are two IETS steps and the separation between these steps increases with increasing $J$, while for antiferromagnetic coupling ($J > 0$), there are three IETS steps (Fig. S4). None of these simulations are consistent with the experimental results. Here, we exclude the E-term, which would otherwise potentially split the states and lead to additional steps. A similar double-step feature, consistent with experimental dI/dV, can be qualitatively reproduced by using D ~ 20 meV and |J| ~ 1. However, this is not consistent with our other experimental results, such as the dI/dV of the FePc-[Fe(C$_6$H$_6$)]$_2$ complexes

(Supplementary Section 5) data and remote sensing experiments (Supplementary Section 4).

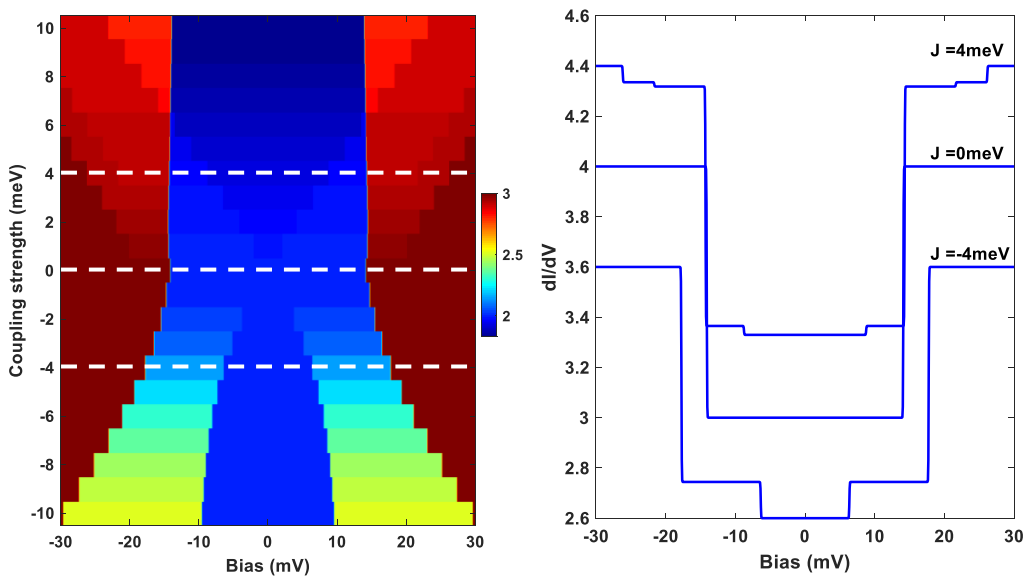

**Supplementary Fig. S4. Simulated dI/dV spectra of a coupled spin system consisting of a S=1/2 and a S=2.** For the $S = 2$ system, we assume anisotropy parameters $D = -4.7 \text{ meV}$, $E = 0 \text{ meV}$. Left: Colormap of simulated spectra atop the $S = 2$ system using different coupling strengths from $J = -10 \text{ meV}$ to $10 \text{ meV}$. Right: Selected spectra for $J = -4 \text{ meV}$, $0 \text{ meV}$ and $4 \text{ meV}$. The spectra are shifted with an offset for clarity. Simulation uses 0.2 K and 0 T.

## 2) FePc S = 1, Fe(C$_6$H$_6$) S = 2

In this case, we assume a spin of 1 for FePc, as it does in the gas phase, and that Fe retains S = 2. Moreover, we assume Heisenberg exchange coupling between them. An exemplary choice of different anisotropy and coupling parameters is displayed in Fig. S5 for coupling strength J ranging from -10 meV to 10 meV. We find that one can adjust J, D, and E-term to reproduce the double steps at around $\pm 20 \text{ meV}$ (Fig. S5), but there are more IETS steps at larger energies and always one step at around zero bias, which is absent in our experimental data.

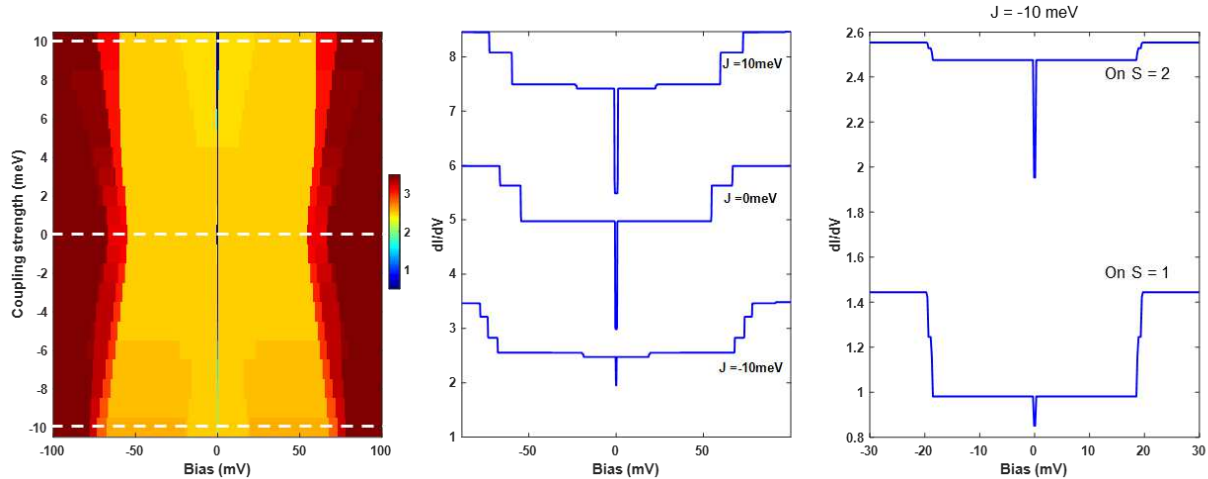

**Supplementary Fig. S5. Simulated dI/dV spectra of a coupled spin system consisting of a S=1 and a S=2.** Anisotropy parameters: $D = -20$ meV, $E = 2$ meV (for $S = 2$) and $D = -3$ meV, $E = 2$ meV (for $S = 1$) were exemplarily chosen, but a wide set of parameters was tested. Left: Colormap of simulated spectra for the $S = 2$ system using different coupling strengths from $J = -10$ meV to $10$ meV. Middle: Selected spectra for $J = -10$ meV, $0$ meV and $+10$ meV. Right: Zoom in plot for both FePc and Fe with $J = -10$ meV, showing two IETS steps at around $\pm 20$ meV. The spectra are shifted with an offset for clarity. Simulation uses 0.2 K and 0 T.

Besides the cases illustrated here, we also tested for instance a spin S = 1/2 coupled to a spin S= 3/2 or 5/2, which always results in more than two IETS steps or of two steps with incorrect position or intensity. Further constraints on the spin system will be discussed in section S4 and S5, which is only fulfilled by the spin model proposed and discussed at the beginning of this chapter.

### 4. Magnetic sensing

To gain further insight into the spin state and magnetic moment of the Fe(C$_6$H$_6$) in the complex, we used a close-by single Fe atom as a magnetic sensor to sense the complexes magnetic field[7]. The magnetic dipole–dipole interaction between the sensor Fe atom and the target spin causes a change of the resonance frequency of the sensor Fe, that is shifted to higher or lower frequencies depending on the spin direction and whether it is in its ground or excited state. Due to its large magnetic anisotropy, the magnetic moment $\mu_{\text{Fe}}$ of the sensor Fe is oriented out-of-plane along the applied magnetic field. Therefore, only the z-component of the target magnetic moment $\mu_t^z$ is measured and the frequency difference $\Delta f$ between its ground and excited state is given by[7]

$$\Delta f = \frac{4E_{dd}}{h} = \frac{1}{h} \cdot \frac{\mu_0}{\pi} \frac{1}{r^3} \mu_{Fe} \, \mu_t^z \qquad (15)$$

where $h$ is Planck's constant, $\mu_0$ is the vacuum permeability and $r$ is the distance between the two magnetic moments $\mu_{Fe}$ and $\mu_t$. In the experiment, the ESR signals are measured using tip-field sweeps[8,9].

As a reference, we first measured the ESR signal of a single Fe atom that shows a single resonance peak (Fig. S6, top). Second, we investigated an Fe-Fe dimer with a distance of 0.91 nm. The splitting of the resonance peak is $\Delta f = (2.1 \pm 0.3)$ GHz (Fig. S6, middle). From this, the magnetic moment of Fe is determined to be $\mu_{Fe} = (5.6 \pm 0.4) \, \mu_B$, which agrees well with previous measurements of $\mu_{Fe} = 5.4 \, \mu_B$[7]. This good agreement indicates that magnetic dipole-dipole interaction is dominant at this distance, despite the close distance between the two spins - below 1 nm. Here, exchange interaction can often contribute significantly to the coupling[7].

Next, we position a sensor Fe atom 0.91 nm from the Fe site of the complex, the same distance as in the previous experiment. The influence of the FePc spin is negligible as its distance to the sensor Fe is roughly twice that of the Fe($C_6H_6$) in the complex. We find a frequency splitting of $\Delta f = (0.9\pm0.2)$ GHz, which is roughly half of the frequency splitting observed for a single Fe atom at the same distance. This corresponds to a magnetic moment of $\mu_{Fe(C6H6)} = (2.8 \pm 0.8) \, \mu_B$. This significantly smaller magnetic moment is incompatible with a $S = 2$ of the Fe atom. While the value of 2 $\mu_B$, indicative of a spin 1 system without orbital contributions, is still within the error bar, we believe that likely exchange interaction is increasing the splitting due to the small distance between sensor and target[2,7]. This subsequently leads to a higher apparent magnetic moment in the sensing experiments.

Consequently, this reduction strongly suggests that the spin state of Fe($C_6H_6$) in the complexes is lower than S = 2.

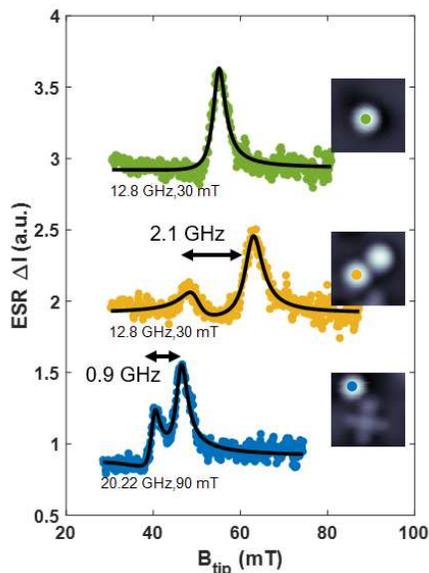

**Supplementary Fig. S6. Magnetic sensing experiment.** Tip-field sweep ESR measurements on a single Fe (green), a sensor Fe atom close to another Fe (yellow) and a sensor Fe atom close to the Fe($C_6H_6$) of an FePc-Fe($C_6H_6$) complex (blue). All data was taken with the same magnetic microtip. Insets: topographic images (size: 2 nm × 2 nm). The inserted numbers display the used external field and RF frequency for the tip-field sweep. The resonance peaks are split by 2.1 GHz (yellow) and 0.9 GHz (blue), respectively. A linear background was subtracted from the data stemming from a general increase in signal with increase of the setpoint current. The black line indicates fits to one or two Lorentz functions.

## 5. dI/dV of an FePc-[Fe($C_6H_6$)]$_2$ complex

A consistency check for our spin model of the FePc-Fe($C_6H_6$) complex is to double the number of Fe atoms in the complex. Figure. S7a (see also Fig. S2c) shows an FePc-[Fe($C_6H_6$)]$_2$ complex, constructed via atom manipulation, where the FePc molecule hosts two Fe($C_6H_6$) complexes. The dI/dV measurements on FePc-[Fe($C_6H_6$)]$_2$ complexes show a pronounced IETS step at around $\pm 40$ meV, instead of around $\pm 20$ meV observed in FePc-Fe($C_6H_6$) complexes. This suggests that the strength of the exchange interaction is simply the same, but added due to the increased number of Fe atoms. Thus, we employ a Hamiltonian

$$H = \sum_{i=1}^{2} J \cdot \vec{S}_{\text{FePc}} \cdot \vec{S}_{\text{Fe(C6H6)},i} + D \cdot S_{z,\text{Fe(C6H6)},i}^{2} \qquad (16)$$

Mainly, an additional exchange interaction is included between the second Fe(C$_6$H$_6$) and FePc in the Hamiltonian. Both $S_{\text{Fe(C6H6)},1} = S_{\text{Fe(C6H6)},2} = 1$. The results of the spin transport calculations are shown in Fig. S7b and reproduce the dI/dV spectra, in which we use $J = 15$ meV and $D = 2.5$ meV, slightly adjusting those values derived from the FePc-Fe(C$_6$H$_6$) complex ($J = 14.65$ meV and $D = 1.9$ meV). The slight non-monotonous background in the spectra are likely caused by many-body effects, which can be accounted for by higher order perturbation theory[5], but have been excluded here for the sake of simplicity. We find additional IETS steps in the dI/dV spectra corresponding to transitions between additional energy levels in the FePc-[Fe(C$_6$H$_6$)]$_2$ complexes. We obtained similar spectra for FePc-[Fe(C$_6$H$_6$)]$_2$ complexes in which the two Fe atoms are located on opposite ligand sites of FePc. We take the consistency of our spin model with the FePc-[Fe(C$_6$H$_6$)]$_2$ complex spectra as further evidence for the validity of the spin model, in particular that the Fe spin state in the complexes is S = 1.

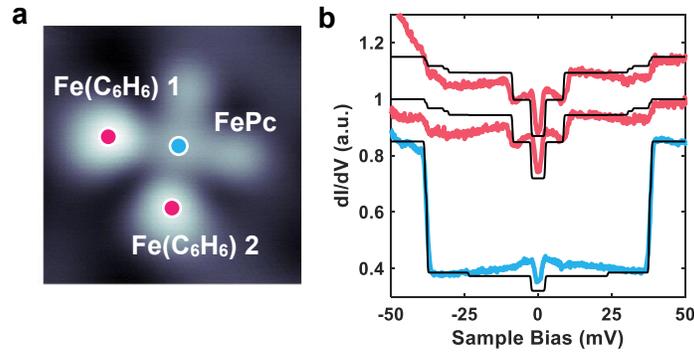

**Supplementary Fig. S7. FePc-[Fe(C$_6$H$_6$)]$_2$ complexes and their spin structure. a,** Topographic image of an FePc-[Fe(C$_6$H$_6$)]$_2$ complex (2nm × 2 nm, setpoint: *V* = 100 mV, *I* = 20 pA). **b,** dI/dV spectra acquired at FePc center (blue curve) and two Fe sites (red curve) as indicated in a. Black lines show the results obtained from the spin transport model and the spin Hamiltonian described in the text. Fitting parameters: For Fe(C$_6$H$_6$): D = 2.5 meV, U = 0.75. For FePc: U = 0.35. J = 15 meV, T=0.8 K.

## 6. DFT calculations

### 1) Details of DFT parameters

All periodic-cell density functional theory (DFT) calculations were performed using plane-wave basis and pseudopotentials in Quantum Espresso version 7.1[10,11]. All pseudopotentials use the Perdew–Burke–Ernzerhof (PBE) parametrization for the exchange and correlation potential[12]. We mimic the experiment by placing the Fe on top of MgO (100) which is supported by a silver (100) substrate. The bulk lattice constants for silver and MgO with PBE are $a_{Ag}$ = 4.16 Å and $a_{MgO}$ = 4.25 Å, which results in a lattice mismatch of about 2% (experimental value: 2.9%[12]). To construct the surface slab, we used 4 layers of Ag with the lateral lattice constant fixed to that of the PBE bulk silver and added up to 2 layers of MgO. We created lateral supercells of about 20x20 Å and 15 Å of vacuum were used to pad the cell in z-direction. All calculations used a k-grid equivalent to 15x15x1 k-points of the 1x1x1 unit cell. All 3d plots in the main text were created using OVITO ref. [13].

## 2) A Fe($C_6H_6$) toy model

To qualitatively understand the effect of Fe with an FePc ligand ring atop, we employ a toy model by placing a benzene ring on top of the Fe atom (as in Fig. 1d/1e but without MgO). Our DFT calculations show that without the presence of a benzene ring, the Fe *d*-states of individual Fe atoms are mostly found close to the Fermi level. For Fe with benzene ring atop, Fe($C_6H_6$), the crystal field generated by the benzene ring leads to a change of the orbital order and spin state that leads to bonding between Fe and ($C_6H_6$) and new molecular orbitals. Combining simple molecular orbital theory with our DFT (Fig. S8), we find a strong overlap between the benzene E1 orbitals and $d_{xz}$, $d_{yz}$ of the Fe, forming a strong set of π-bonds. This is also commonly the case for other metallocenes[14] and also rationalizes the stability of the Fe($C_6H_6$) complex. Similarly, the frontier orbitals are formed by an overlap between $d_{xz}$, $d_{yz}$ and the ($C_6H_6$) E1 antibonding states. Still, the frontier orbitals maintain a strong *d*-character (Fig. S9a). For instance, the $d_{xz}$-, $d_{yz}$-like orbitals account for 76% of $2e_1$ near $E_F$.

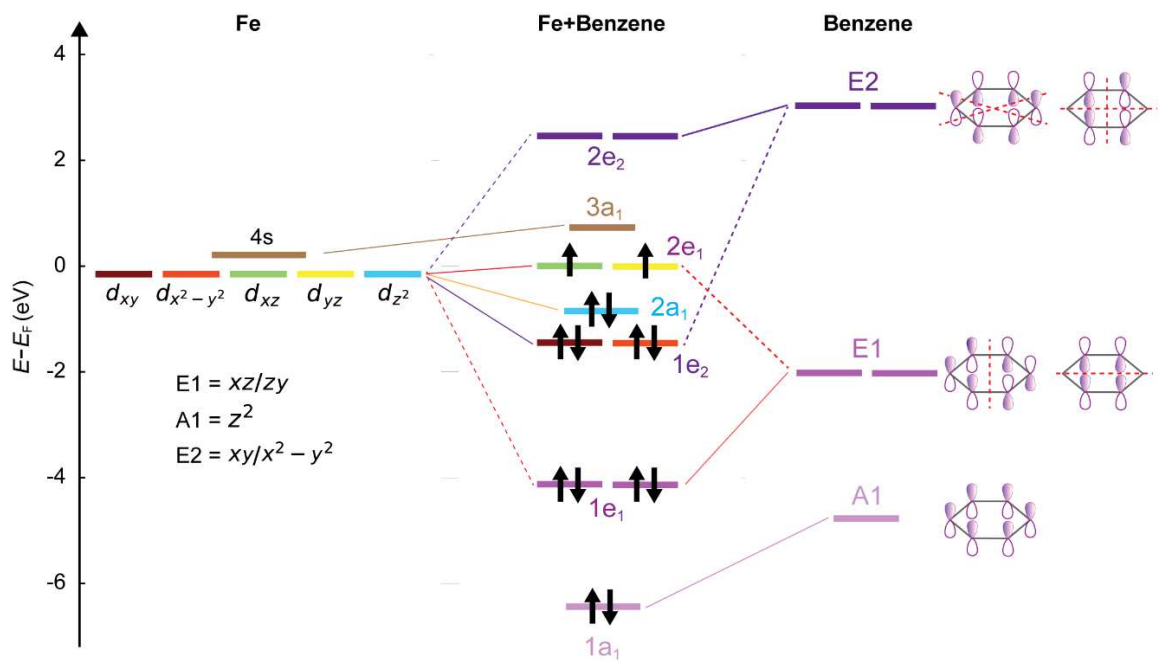

**Supplementary Fig. S8. Molecular orbital energy diagram of Fe and Benzene (in vacuum).** On the left side, Fe 3*d* and 4*s* states are shown with their respective symmetry group, on the right side the $\pi$ molecular orbitals of benzene. In the center the combined molecular orbitals of Fe(C$_6$H$_6$) are shown. The lowest lying states are dominated by the benzene states (1a$_1$ 90%; 1e$_1$ 84%) with some overlap of the 3*d* and 4*s* states. For the higher ones, the main contribution stems from the *d* states (1e$_2$ 72%; 2a$_1$ 93%; 2e$_1$ 76%). Energies and orbital character are obtained from DFT calculations. The addition of the MgO/Ag(001) surfaces leads to further hybridization of all states (See supplementary Fig. S9).

## 3) DFT calculations of the full system

We first confirmed that DFT accurately reproduced S = 2 of individual Fe atoms on the MgO/Ag(100) surface as discussed elsewhere[15]. From the local density of states (LDOS) in Fig. 1d, the Fe *d*-states are mostly found close to the Fermi level with the $d_{z^2}$ orbital strongly hybridizing with the 4*s* orbital. This hybridization results in the $d_{z^2}$ orbital being half-filled due to on-site repulsion. As a consequence, Fe has a final electron occupation of $3d^{6.5}4s^{0.8}$ obtained from a Lowdin charge analysis and four orbitals ($d_{z^2}/d_{xy}/d_{xz}/d_{yz}$) are close to being half-filled (Supplementary Table 1). This consequently indicates a spin $S = 2$, consistent with previous works[1,15]. To understand how the Fe(C$_6$H$_6$) spin state in the complex becomes $S = 1$, we employ again the simple model by placing a benzene ring on top of the Fe atom (Fig. 1e). We find that the benzene ring does not change the charge state of the Fe atom: Here, Fe exhibits a $3d^{6.8}4s^{0.3}$ configuration, indicating no significant charge transfer (Supplementary Table 1). However, the crystal field generated by the benzene ring changes the $d$ orbital order: the $d_{z2}$, $d_{xy}$ and $d_{x2-y2}$ like orbitals are now shifted and lowered in energy, and the $4s - 3d_{z^2}$ hybridization is eliminated due to the depletion of the 4s in the presence of benzene ring. Consequently, only the $d_{xz}/d_{yz}$ like orbitals are close to the Fermi energy and half-filled, suggesting an Fe spin state of $S = 1$.

| Configuration | $d_{z2}$ | $d_{xy}$ | $d_{xz}$ | $d_{yz}$ | $d_{x2-y2}$ | total |
|---|---|---|---|---|---|---|
| Fe (↑) | 0.9263 | 0.9965 | 0.9962 | 0.9962 | 0.9836 | **4.9** |
| Fe (↓) | 0.5274 | 0.0034 | 0.0472 | 0.0472 | 0.9628 | **1.6** |
| Fe+benzene (↑) | 0.8525 | 0.8542 | 0.959 | 0.9592 | 0.8386 | **4.5** |
| Fe+benzene (↓) | 0.6793 | 0.6932 | 0.1314 | 0.1323 | 0.6826 | **2.3** |

**Supplementary Table 1**. Filling of the 3d shell obtained from a Lowdin charge analysis for Fe on 2ML MgO/Ag(001) without and with a benzene ring on top.

To shed light on the role of the substrate, we compare Fe-benzene formation in vacuum and on the MgO/Ag(001) surface (Figure S9). While there is more overlap with other states, the simple molecular orbital picture derived from the toy model still remains valid: Figure S9 shows that both deliver similar results. However, the surface decreases the energy splitting between $d$ states and further mixes the states by hybridization with the substrate.

DFT calculations including the full FePc molecule do not change our observations (see Fig. S10). The 3$d$ states near the Fermi level show a similar trend to Fe(C$_6$H$_6$), however the tilted final geometry makes the analysis of the orbital character of the Fe 3$d$ states less straightforward.

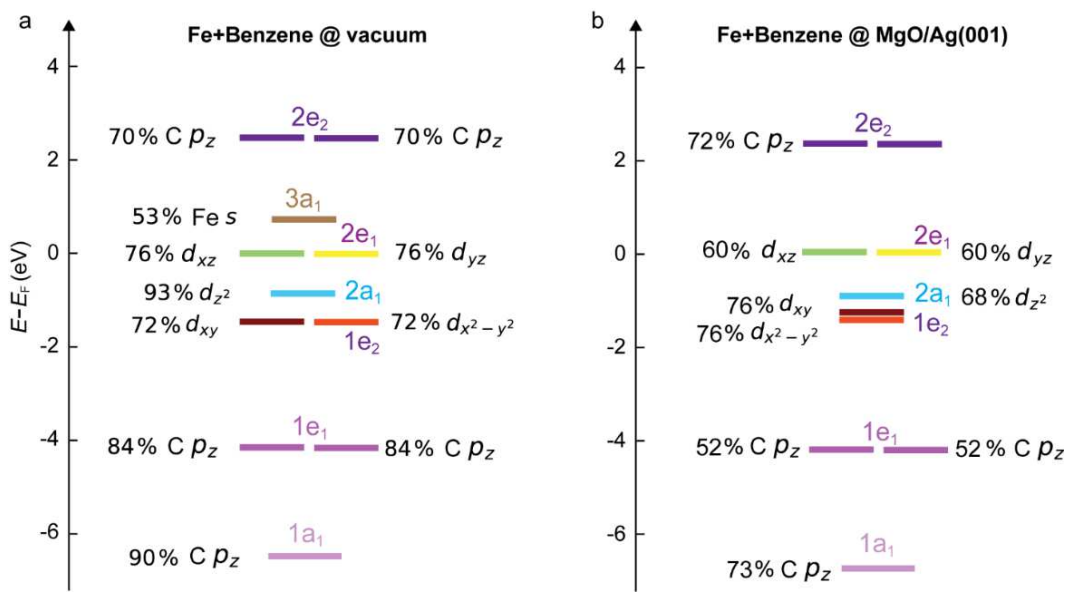

**Figure S9. Molecular orbital energy diagrams of Fe(C$_6$H$_6$) in different environments. a,** in vacuum and **b,** on a MgO/Ag(001) substrate. The orbital weights are given in percent (%) listing the orbital of the dominating weight. The remaining contributions to 100% are hybridizations with other orbitals. Energies and character are obtained from DFT calculations.

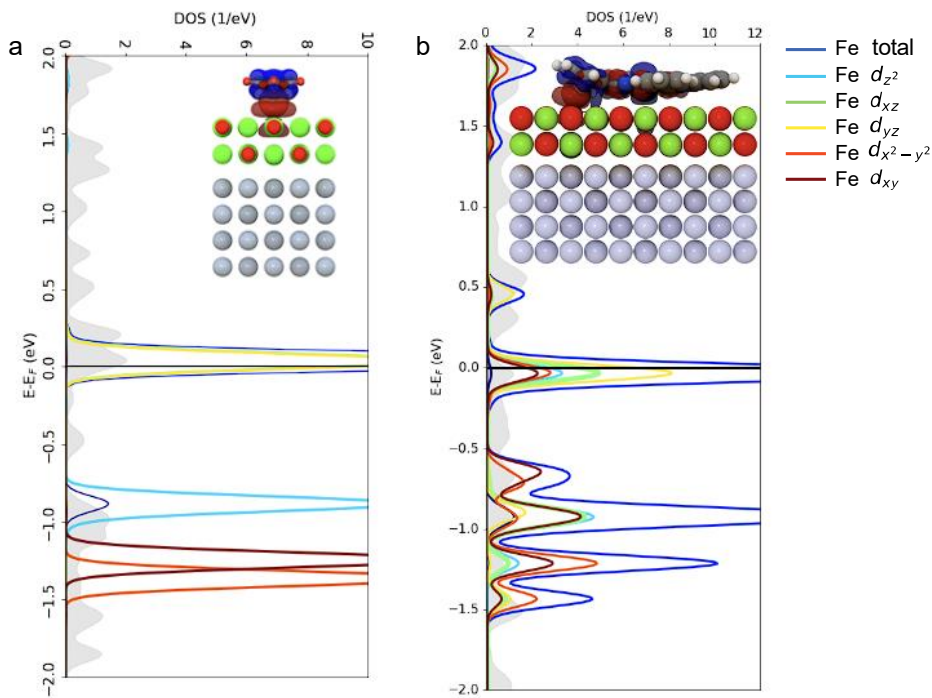

**Supplementary Fig. S10. Atomic orbital projected DOS of Fe(C$_6$H$_6$) and FePc-Fe(C$_6$H$_6$). a,** Fe 3$d$ orbitals in the Fe-benzene complex used in the main text. **b,** the same plot but now with an FePc-Fe(C$_6$H$_6$) complex. The resulting states show more mixing but the behavior of the states close to $E_F$ follows a similar trend as in a.

## 4) Exchange interaction within the complex

We use DFT calculations of an FePc-Fe(C$_6$H$_6$) complex to estimate the exchange coupling between the two spin centers using the broken-symmetry approach[2,16]. The magnetic exchange coupling strength is obtained by comparing the energy of the high spin (HS) and low-spin or broken symmetry (BS) states as:

$$J = \frac{-[E(HS) - E(BS)]}{s_{max}^2}$$

The high-spin solution corresponds to FM coupling whilst the BS solution is obtained from an antiferromagnetically coupled configuration and $s_{max}$ is the total spin of the high-spin state. We obtain an exchange coupling $J$ for the Fe-FePc complexes of around 7 meV, which is of the same order as the experimental results (14 meV). We emphasize that this is the largest $J$ observed for spins on MgO and is mediated via the FePc ligand and by far exceeds the exchange coupling of two Fe atoms at the same distance but without the ligand. This indicates that the exchange coupling is mediated through the ligand as also observed in FePc-FePc dimers[2] and FePc-Ti dimes[17]. We confirm that the exchange coupling in this system is strongly mediated by the ligand by slowly lifting the ligand up from the Fe atom in the calculation (Fig. S11). In such a model, the exchange coupling constant is approximately exponentially dependent on the Fe atom – ligand distance. As the distance increases by tilting the ligand from 0° to 20°, the exchange coupling J decreases exponentially from 7 meV to 0.29 meV.

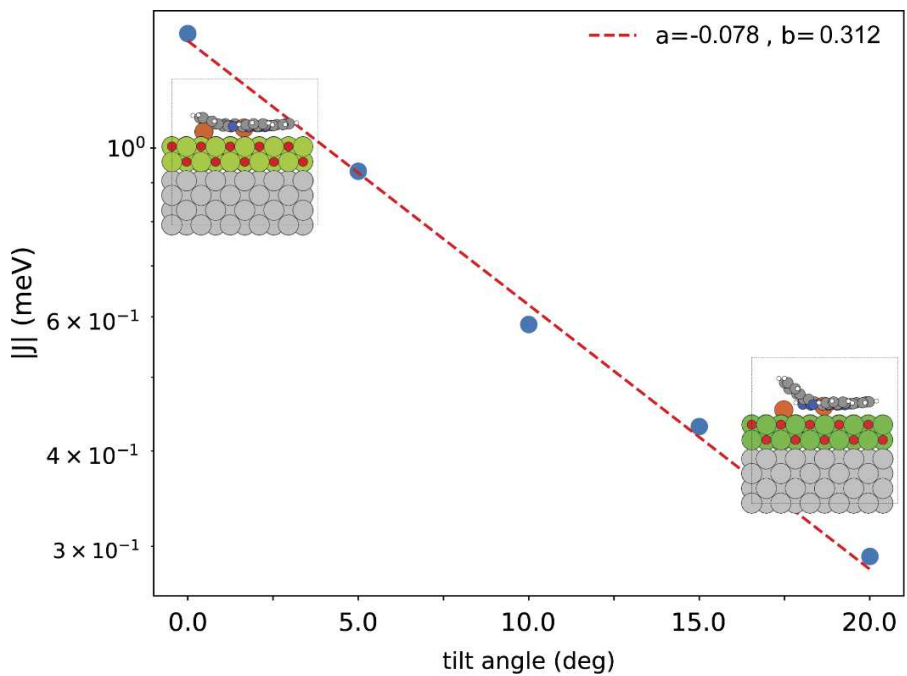

**Supplementary Fig. S11. Exchange energy as a function of Fe atom – ligand distance.** Inset: sketch showing 0° (left) and 20°(right) tilt angle configurations. The red dash line is a linear fit to data in log scale.

## 7. ESR and fitting

Figure 2(e-f) in the main text show the external magnetic field dependent ESR measurement at the FePc site and Fe($C_6H_6$) site of the same FePc-Fe($C_6H_6$) complex with the same tip. Fitting $f_0$ linearly as a function of $B_z$ gives the magnetic moment of (1.008±0.007) $\mu_B$ at the Fe($C_6H_6$) site (Fig. S12a) and (1.004±0.012) $\mu_B$ at the FePc center (Fig. S12b), which is consistent to the effective spin ½ ground state doublet. Figure S12c shows the magnetic moments at both sites obtained for different FePc-Fe($C_6H_6$) complexes, giving an averaged magnetic moment of $(1.003 \pm 0.025)\mu_B$. While this is close to an effective magnetic moment of 1 $\mu_B$, the variations are larger than the individual errorbars. We suggest that this results from the influence of the magnetic tip field[18] as well as changes in the local environment of the spin center, i.e. given by defects and strain in the substrate.

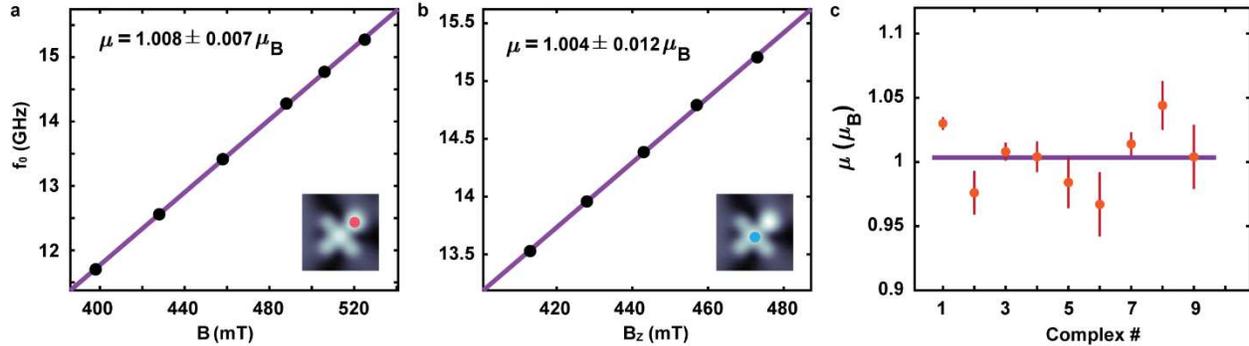

**Supplementary Fig. S12. Magnetic moment of FePc-Fe(C$_6$H$_6$) complexes.** Resonance frequency $f_0$ as a function of an out-of-plane external magnetic field $B$. **a,** Fe(C$_6$H$_6$) site ($I_{set}$ = 5 pA / $V_{set}$ = 25 mV); **b,** FePc site ($I_{set}$ = 50 pA / $V_{set}$ = 70 mV) extracted from Fig. 2e,f in the main text. **c,** Statistics of magnetic moments in different FePc-Fe(C$_6$H$_6$) complexes.

## 8. Rabi rate as a function of $V_{RF}$

We extract the Rabi rate at different $V_{RF}$ from Fig. 3b in the main text, and a linear fit yields a slope of $9.6 \pm 0.3 \text{ rad}/(\mu s \cdot mV)$ (Fig. S13). This slope is 5 times higher than that of pristine FePc [$1.86 \pm 0.13 \text{ rad}/(\mu s \cdot mV)$ in Ref.[19]] and 3-7 times larger than values reported for Ti atoms [$\sim 1.3 \text{ rad}/(\mu s \cdot mV)$ in Ref.[20] and $\sim 2.8 \text{ rad}/(\mu s \cdot mV)$ in Ref.[21]]. In the following we discuss possible reasons for these results.

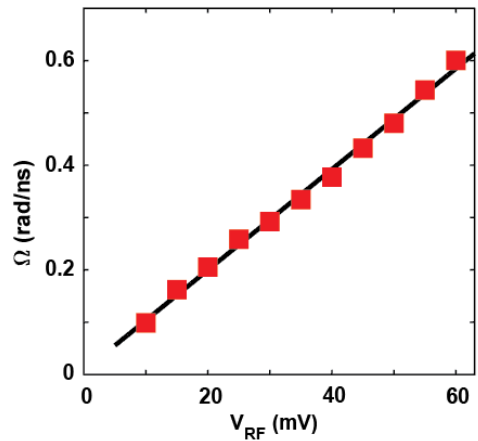

**Supplementary Fig. S13.** Extracted Rabi rate $\Omega$ as a function of $V_{RF}$ from Fig. 3b in main text. The black line is a linear fit to the data with a slope of $9.6 \pm 0.3 \text{ rad}/(\mu s \cdot mV)$ .

The driving mechanism of individual atoms and molecules in ESR-STM is still subject to current research, both theoretically and experimentally[21-25]. We first discuss the ESR driving mechanism in the framework of the frequently employed piezoelectric displacement model[22], which relies on the displacement of the surface spin and a subsequent modulation of one of the parts of the Hamiltonian. Employing the same formalism as presented in Ref. [22], we write

$$\Omega(t) = \frac{F}{h} z(t)$$

where the Rabi rate $\Omega$ depends on the oscillating displacement $z(t)$ and the Rabi force $F$. The latter can have different origins[22]. The dominating term was found theoretically[22] and experimentally[20] to originate from exchange interaction of the tip magnetic moment and the surface spin leading to an effectively oscillating local magnetic field. Thus, a higher Rabi rate can originate either from an increase in $z(t)$ or $F$.

A straightforward description of the displacement is given by

$$z(t) = \frac{q}{k} \frac{V_{RF}(t)}{d}$$

Here, it is assumed that the time-dependent electric field $E(t) = \frac{V_{RF}(t)}{d}$ couples via Hooke's law to the surface spin with an effective charge $q$ and effective spring constant $k$, i.e. the stiffness of the vibrational mode of the spin center on the surface.

Thus, a larger displacement can either be caused by a different $q$ of the surface spin, a less stiff interaction to the surface $k$, or simply a different initial distance $d$ between the surface spin and the tip.

In addition, the Rabi force itself is potentially stronger for the Fe(C$_6$H$_6$) site in the complex:

$$F = \frac{\partial J(z)}{\partial z} \langle \vec{S}_T \rangle \cdot \langle 0|\vec{S}|1 \rangle$$

Here, $\frac{\partial J(z)}{\partial z}$ is the gradient of the exchange interaction between the surface spin and the tip magnetic moment $\langle \vec{S}_T \rangle$. $\langle 0|\vec{S}|1 \rangle$ is the transition matrix element, emphasizing that a finite overlap between two states is necessary for their connection. Thus, both $\frac{\partial J(z)}{\partial z}$ and $\langle 0|\vec{S}|1 \rangle$ can cause an increase in $\Omega$. While the former is difficult to estimate, the latter can be calculated from the Hamiltonian of the FePc-Fe(C$_6$H$_6$) complex given in the main text for the Fe(C$_6$H$_6$) site. When comparing this to a normal spin ½, we find a 33% increase (see Supplementary Table 2). This increase is very likely to play a role in the faster Rabi rate observed for the FePc-Fe(C$_6$H$_6$) complex, but cannot explain the full effect.

Beyond the piezoelectric displacement model, it was proposed by some of the authors[24], that the electric field of the tip can directly modulate the charges of the atoms without displacement. This leads for spin dimers to a modulation of the J-coupling and thus explains experimental work on Fe-Ti spin pairs quite well[21]. Thus, for the FePc-Fe(C$_6$H$_6$) complex, J-modulation between the Fe and FePc spin adds another possibility for a higher Rabi rate.

Lastly, in another recent work, some of the authors proposed an electron transport model[26], which predicts that the magnitude of the Rabi term mainly depends on tip polarization and the DC bias magnitude compared to the ionization energy of the adsorbate. In particular, the latter changes for the new molecular orbitals emerging in the complex (Fig. 1e).

In general, we believe that a combination of the different origins discussed here play a role for the enhanced Rabi rate: This includes the summarized effects to the displacement $z(t)$, differences in the Rabi force, the strong exchange coupling as well as a change in ionization energy in the dimer.

| Matrix Element | Spin ½ (FePc) | Spin (1, ½) Ferrimagnet (Complex) |
|---|---|---|
| $|\langle 0|S_x|1\rangle|$ | 0.5 | 0.667 |
| $|\langle 0|S_y|1\rangle|$ | 0.5 | 0.667 |
| $|\langle 0|S_z|1\rangle|$ | 0 | 0 |

**Supplementary Table 2.** Table of transition matrix elements of the respective ESR transition for different spin systems. For the FePc-Fe(C$_6$H$_6$) complex, the transition matrix element is determined for the Fe site which corresponds to the case in the Rabi measurements.

## 9. Conductance-dependent spin lifetime

To further investigate the processes that limit the spin lifetime $T_1$, we performed pump-probe measurements as a function of tip-sample conductance, $G_{ts} = I/V$, for both pristine FePc (Fig. S14a) and FePc in the complex (Fig. S14b). In both cases, the lifetime drops dramatically as the tip approaches the spin centers. At large tip-sample distances (i.e. low $G_{ts}$), $T_1$ reaches a plateau where it becomes independent of tip-sample distance. This plateau, defined as $T_1^0$, is also independent of the specific tip and molecules in use. We therefore conclude that $T_1^0$ represents the intrinsic relaxation lifetime on a 2ML MgO surface. The effect of the tip height on $T_1$ can be understood by a model based on electron-hole generation in the nearby electrodes, i.e. the metallic substrate and tip, as described in Ref.[27] for single Fe atoms on MgO/Ag(001). According to this model, $T_1$ is a function of sample–sample ($G_{ss}$), tip–sample ($G_{ts}$) and tip–tip ($G_{tt} = \frac{G_{ts}^2}{G_{ss}}$) conductance: $T_1 = \frac{T_1^0}{(1+\frac{G_{ts}}{G_{ss}})^2}$ (Fig. S14c). Fitting $T_1$ as a function of $G_{ts}$ yields $T_1^0$ of $(0.41 \pm 0.05)$ μs for pristine FePc (solid blue curve in Fig. S14a) and $(1.6 \pm 0.6)$ μs for the complex (solid red curve in Fig. S14b). In Figure S14d, we show the longest $T_1$ achieved at a large tip-sample distance ($G_{ts} = \frac{20}{200}$ pA/mV). The data reveals a $T_1$ of $(1428 \pm 323)$ ns for FePc in the complex and $(363 \pm 26)$ ns for pristine FePc, which are consistent with the values from our model. At large tip-sample distance, the lifetime is limited by the sample-sample conductance $G_{ss}$. We believe that $T_1$ can be further enhanced by increasing the MgO layer thickness as it is the case for individual Fe atoms[27].

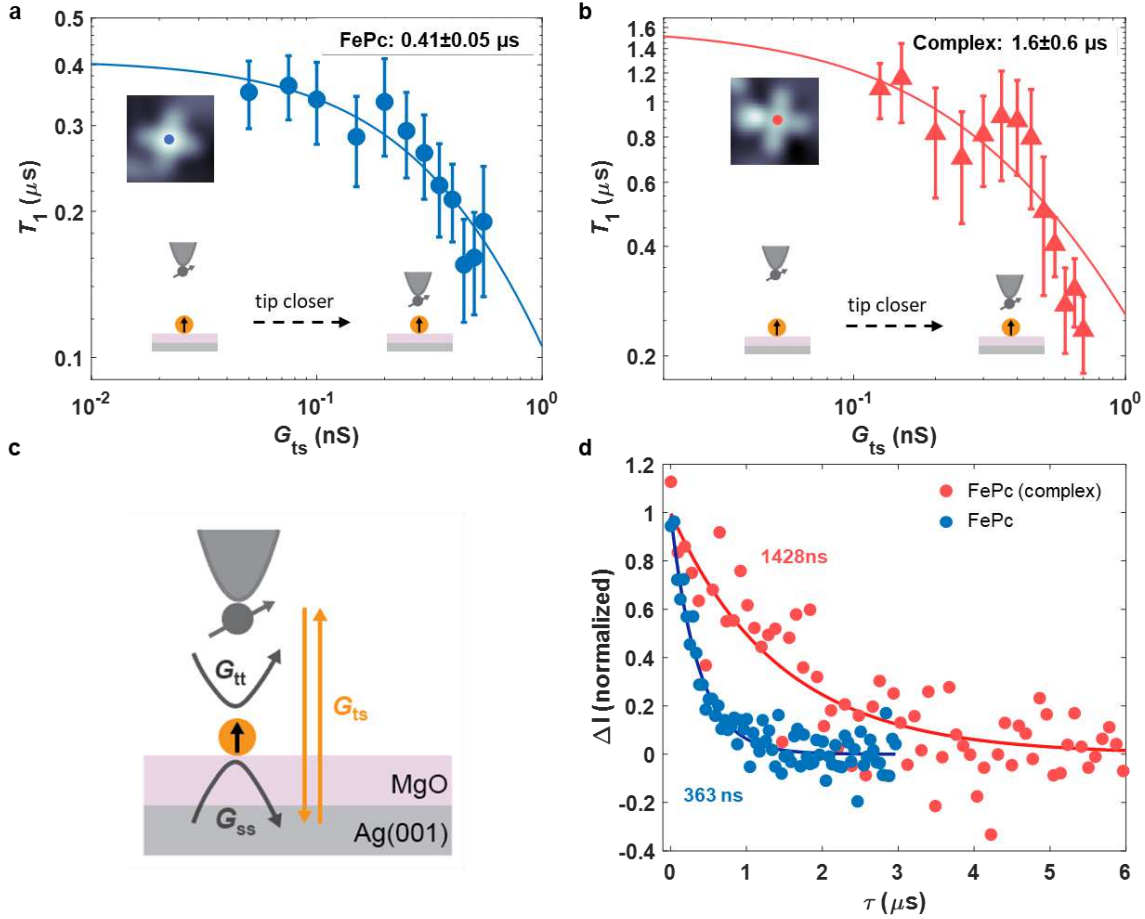

**Supplementary Fig. S14. Conductance-dependent spin lifetime $T_1$ as a function of tip-surface conductance**. **a,b** $T_1$ as a function of tip-surface conductance for pristine FePc and FePc site in the complex. (setpoint: $V_{set} = 200$ mV, $I_{set} = = 10 - 100$ pA for FePc, $I_{set} = 25 - 140$ pA for the complex). All data are measured with the same tip. Error bars are determined by exponential fits to pump–probe data. The solid lines are fits. Fitted parameters: $T_1^0 = (1.6 \pm 0.6)$ μs, $G_{ss} = 0.4$ nS (complex), and $T_1^0 = (0.41 \pm 0.05)$ μs, $G_{ss} = 0.21$ nS (FePc). **c**, Illustration of tip–tip ($G_{tt}$), tip–sample ($G_{ts}$) and sample–sample ($G_{ss}$) conductance paths. Their relative strength can be tuned by adjusting the distance of the metallic tip to the spin. $G_{tt} = G_{ts}^2/G_{ss}$. **d**, $T_1$ for a pristine FePc (blue) and FePc in the complex (red) on 2 ML MgO/Ag(001) at low $G_{ts}$. [parameters: $I_{set} = 20$ pA, $V_{set} = 200$ mV, $B = 600$ mT, $V_{pump} = 80$ mV, $V_{probe} = 40$ mV, $\tau_{cycle} = 6$ μs (FePc site in the complex), $\tau_{cycle} = 3$ μs (FePc)].

## 10. Spin transport simulations and enhanced spin lifetime

### 1. Spin transport simulations

In the following, we discuss the spin transport simulation in greater detail. These are employed to explain the improvement in spin lifetime, which in the main text, we ascribe to the correlation of the two spins in the complex. For a electron tunneling process through an STM tunnel barrier featuring a single spin, the intensity of the transition from a given initial state $|\varphi_i, \psi_i\rangle$ to a final state $|\varphi_f, \psi_f\rangle$, i.e. the transition matrix element, is given as (Eq. 13 and 14 in Ref.[5])

$$|M_{if}|^2 = \left|\langle\varphi_f, \psi_f|\tfrac{1}{2}\mathbf{S}\cdot\boldsymbol{\sigma} + U|\varphi_i, \psi_i\rangle\right|^2 = |m_{if}|^2 + |U|^2\delta_{if} + 2\cdot\text{Re}[u \times m_{if}]\delta_{if} \qquad (15)$$

Here, $|\varphi\rangle$ ($|\psi_i\rangle$) describes states of the electron baths (spin system). The interaction between the tunneling electron and the localized spin is described as an exchange interaction of the form $\tfrac{1}{2}\mathbf{S}\cdot\boldsymbol{\sigma}$. Here, $\mathbf{S}$ and $\boldsymbol{\sigma}$ are the spin vector operators of the local spin and the tunneling electron, respectively. The (dimensionless) parameter $U$ accounts for spin-independent components of the interaction, i.e. Coulomb potential scattering. $m_{if} = \left|\langle\varphi_f, \psi_f|\tfrac{1}{2}\mathbf{S}\cdot\boldsymbol{\sigma}|\varphi_i, \psi_i\rangle\right|$ also contains the inelastic scattering between the two spins. Fig. S15 illustrates the composition of the different contributions to a d$I$/d$V$ measurement including elastic and inelastic channels as derived in Ref.[5]. From the d$I$/d$V$ measurement, the ratio of inelastic scattering (blue area in Fig. S15) and spin conserving transport processes can be determined in an external magnetic field. In order to obtain the probability for inelastic scattering we thus simulated both FePc and complex spectra and extracted the ratio between inelastic and elastic tunneling channels. We take the spin configuration from the d$I$/d$V$ and ESR analysis performed in the main text (Complex: $S_{\text{Fe(C6H6)}} = 1$, $S_{\text{FePc}} = ½$, $J = 14$ meV, $D = 1.8$ meV, $U_{\text{Fe}} = 1$, $U_{\text{FePc}} = 0.5$; isolated FePc: $S_{\text{FePc}} = ½$, $U_{\text{FePc}} = 0.5$ taken form Ref.[2]). In the simulation, we additionally apply a magnetic field of 5 T in order to separate inelastic and elastic contributions at around zero bias (As shown in Fig. S16a). The exemplary ratio from these simulations are shown in Fig. S16a for both the complex and the pristine FePc. As it can be seen, the probability of inelastic scattering, i.e. the height of the IETS step is significantly reduced for the complex, in particular for the FePc site. In Fig. 3f in the main text we show the continuous evolution of this ratio as a function of exchange coupling between the two spins, which tunes the system continuously from the pristine FePc into the ferrimagnet configuration. Compared to pristine FePc, the reduction in inelastic spin-flip fraction for FePc in the complex is mostly attributed to the large $J$ which leads to strong correlation between two spins, as the U parameter remains similar for them. This emphasizes that the correlation in the spin system is crucial to protect the spin from inelastic scattering contributions in the environment. While these simulations are based solely on the parameters extracted from the zero-field d*I*/d*V* spectra measured with non-spin polarized tips, we can additionally obtain these fractions from experimental d*I*/d*V* spectra which are discussed in the following section.

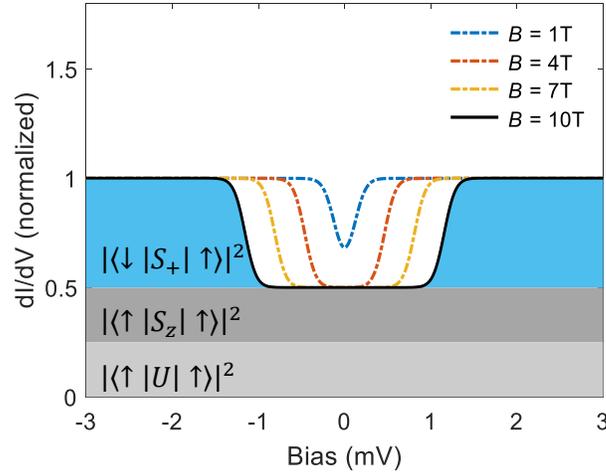

**Supplementary Fig. 15. Simulated tunneling spectra for a S=1/2 system at different magnetic fields.** The different contributions to the conductance are labeled. Blue: Conductance due to inelastic spin-spin scattering with changing the localized spin, Dark gray: elastic spin-spin scattering without changing the spin, light gray. Elastic Coulomb potential scattering with $|U| = 0.25$. The spectra are normalized and g = 2 here. T was set to 0.5 K.

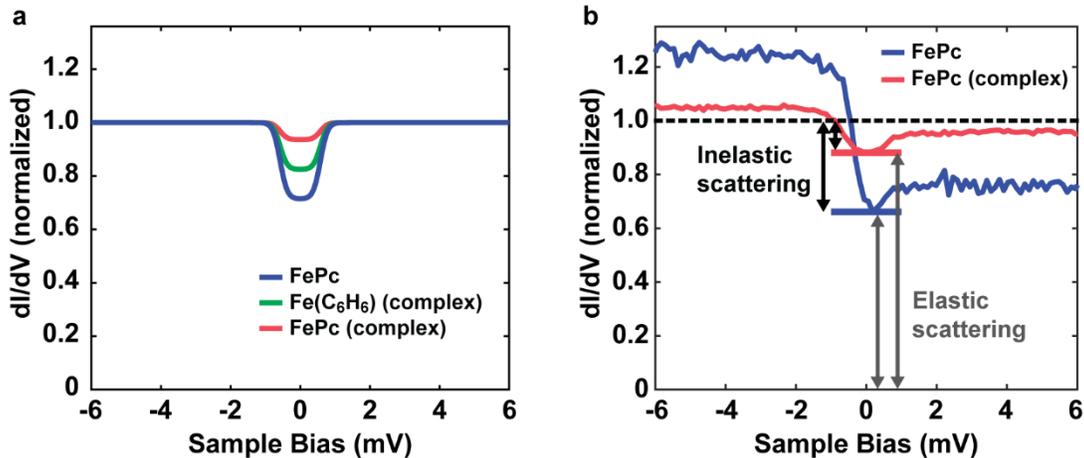

**Supplementary Fig. S16. Inelastic Scattering in d*I*/d*V*. a,** Simulated dI/dV of FePc and of the complex: $S_{Fe(C6H6)} = 1$, $S_{FePc} = ½$, $J = 14$ meV, $D = 1.8$ meV, $U_{Fe} = 1$, $U_{FePc} = 0.5$; isolated FePc: $S_{FePc} = ½$, $U_{FePc} = 0.5$ taken form Ref. [2]. T = 0.8 K, B = 5 T. **b,** Normalized d*I*/d*V* measurements with a spin-polarized tip to determine the fraction of inelastic scattering for the pristine FePc, as well as Fe(C6H6) and FePc in the complex. Data taken with a magnetic tip, in order to induce a large Zeeman splitting (Tunneling Parameter: FePc: $I_{set} = 1$ nA / $V_{set} = -30$ mV; FePc (complex): $I_{set} = 0.5$ nA / $V_{set} = -30$ mV).

## 2. Experimental determination of inelastic scattering probability

Besides the spin transport simulations, which offer a detailed picture of the influence of the exchange coupling and correlation on inelastic scattering processes, we can also experimentally determine the probability of inelastic scattering by performing d*I*/d*V* measurements on the FePc and the complex.

Fig. S16b shows d*I*/d*V* spectra, which were recorded using a magnetic tip with a field strong enough to split the doublet ground state. Usually, an external magnetic field is used for this kind of measurements, which however in our experimental setup is not strong enough to sufficiently split the two states. The spectra in Fig. S16b include both the isolated FePc and measurements on the FePc site in the complex. For all spectra, the signal above the inelastic transition is asymmetric for different bias polarity due to the additional spin-polarization of the magnetic tip[5]. Thus, we normalize the spectra to the average of the two step heights, which corresponds to the level in the non-spin-polarized case. As can be seen, the fraction of inelastic scattering (blue) is largest for FePc, and decreases for the FePc site in the complex. The ratio of inelastic scattering can be read off and is shown for different measurements (different FePcs/complexes and different tips) in the main text in Fig. 3f.

## 11. ESR transitions in a Heisenberg two-spin system with the tip field detuning effect

In the main text, we argue that the two coupled complex spins (four spins) can be treated effectively as two coupled spin ½, such as found in Ref. [2,28]. In the following, we will derive the essential equations necessary to describe the two coupled complexes as effectively two coupled spin ½. Moreover, we will discuss the antiferromagnetically (AFM) coupled case, in contrast to the ferromagnetically (FM) case discussed in the main text.

Using the secular approximation, the Hamiltonian in the main text can be rewritten as

$$H = -g_1\mu_B(B_{tip} + B) \cdot S_1^z - g_2\mu_B B \cdot S_2^z + (J_2^{eff} + 2D) \cdot S_1^z S_2^z + (J_2^{eff} - D)(S_1^x S_2^x + S_1^y S_2^y)$$

Here z is aligned with $\mathbf{B}$, $\mu_B$ is the Bohr magneton, and $J$ and $D$ represent the exchange and dipole coupling strengths, respectively. Labels 1 and 2 indicate the first and second complex. Thus, the first two terms are the Zeeman splittings of the respective complexes. The detuning, i.e. the difference between the Zeeman splittings of the two complex spins, is given by the imbalance of these two terms, i.e. $\delta = g_1 \mu_B (B + B_{\text{tip}}) - g_2 \mu_B B$. The coupling constant for the dipolar coupling is given by $D = \frac{D_0}{2}(1 - 3\cos^2\theta)$ with $D_0 = \frac{\mu_0 \gamma_1 \gamma_2 \hbar^2}{4\pi r^3}$, where $\gamma_{1,2} = g_{1,2} \mu_B / \hbar$ are the gyromagnetic ratios of the two complex spins. Here, $\theta = 90°$ represents the angle between the external field and the connection vector $\mathbf{r}$ of the two complex spins, resulting in $D = \frac{D_0}{2}$. The magnetic dipolar coupling treats the effective coupling between the two complex spins. Using Zeeman product states $|00\rangle, |10\rangle, |01\rangle,$ and $|11\rangle$ as a computational basis, the eigenstates and corresponding eigenenergies of the coupled complex Hamiltonian are:

| n | Eigenstate $|n\rangle$ | Eigenenergy $E_n$ |
|---|---|---|
| 0 | $|00\rangle$ | $-\frac{1}{2}(g_1 \mu_B (B + B_{\text{tip}}) + g_2 \mu_B B) + \frac{1}{4}(J_2^{\text{eff}} + 2D)$ |
| 1 | $\|-\rangle: -\frac{\alpha}{\sqrt{1+\alpha^2}} |01\rangle + \frac{1}{\sqrt{1+\alpha^2}} |10\rangle$ | $-\frac{1}{2}\sqrt{\delta^2 + (J_2^{\text{eff}} - D)^2} - \frac{1}{4}(J_2^{\text{eff}} + 2D)$ |
| 2 | $\|+\rangle: \frac{1}{\sqrt{1+\alpha^2}} |01\rangle + \frac{\alpha}{\sqrt{1+\alpha^2}} |10\rangle$ | $\frac{1}{2}\sqrt{\delta^2 + (J_2^{\text{eff}} - D)^2} - \frac{1}{4}(J_2^{\text{eff}} + 2D)$ |
| 3 | $|11\rangle$ | $\frac{1}{2}(g_1 \mu_B (B + B_{\text{tip}}) + g_2 \mu_B B) + \frac{1}{4}(J_2^{\text{eff}} + 2D)$ |

The coupling terms cause the formation of two mixed states $|+\rangle$ and $|-\rangle$, which are a superposition of the two Zeeman states $|01\rangle$ and $|10\rangle$ and can be characterized by the parameter $\alpha$, which depends on the detuning $\delta$ and the exchange coupling $J_2$.

$$\alpha = \frac{\delta + \sqrt{\delta^2 + (J_2^{\text{eff}} - D)^2}}{J_2^{\text{eff}} - D}$$

$$\delta = g_1\mu_B(B + B_{tip}) - g_2\mu_B B$$

Four possible transitions could occur, labeled by the corresponding ESR frequencies: $f_1$ for $|00\rangle \rightarrow |-\rangle$, $f_2$ for $|+\rangle \rightarrow |11\rangle$, $f_3$ for $|00\rangle \rightarrow |+\rangle$ and $f_4$ for $|-\rangle \rightarrow |11\rangle$. These four transitions are those shown in Fig. 4c-f in the main text and in Figs. S17-19. The frequency difference between $f_1(f_3)$ and $f_2(f_4)$ depends on $J_2^{eff}$ and $D$ given by:

$$\Delta f_1 = f_2 - f_1 = f_4 - f_3 = J_2 + 2D = J_2^{eff} + D_0.$$

When the exchange coupling is dominant ($|J_2^{eff}| \gg D_0$), $\Delta f_1$ can be either positive ($J_2^{eff} > 0$, AFM coupling) or negative ($J_2^{eff} < 0$, FM coupling). In the main text, we demonstrate that the two complex spins are coupled ferromagnetically. However, they can also be coupled antiferromagnetically when the two Fe(C6H6) in each complex are positioned close together. The coupling sign can be determined by comparing the relative ESR peak heights of $f_3$ ($f_1$) and $f_4(f_2)$. Since for the AFM case (Fig. 4 and Fig. S17) the ESR peak at $f_3(f_1)$ coresponds to the ground state of the neighboring complex spin (the spin under the tip), the ESR peak at $f_3$ ($f_1$) is taller than at $f_4(f_2)$. In the FM case, the mixed states $|-\rangle$ and $|+\rangle$ move to higher energies, causing $f_3(f_1)$ to appear at a higher frequency than $f_4(f_2)$. Generally speaking, the peak at lower frequencies is taller for AFM (Fig. S18), while it is at higher frequencies in the FM case (Fig.4).

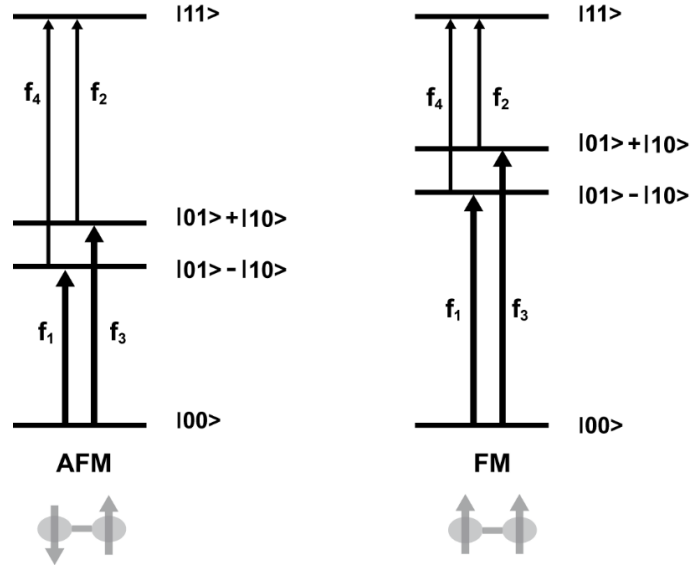

**Supplementary Fig. S17. Energy levels of a pair of AFM or FM coupled spin 1/2.** The arrows indicate the possible ESR transitions. The frequency and intensity are represented by the length and thickness of the arrows. The peak at lower frequencies ($f_1$ or $f_3$) is taller for AFM, while it is higher in the FM case.

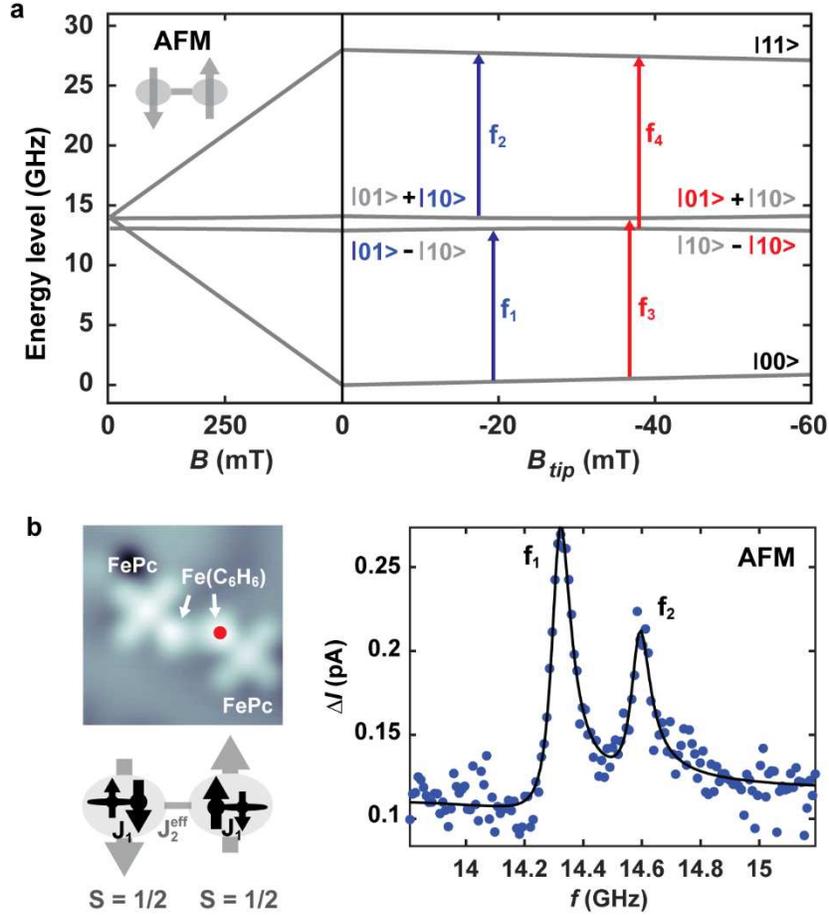

**Supplementary Fig. S18. AFM Coupling in a Dimer of Complexes. a.** Energy level diagram of two AFM coupled spin ½ in the presence of $B$ and $B_{\text{tip}}$. **b.** Topographic image of two coupled complexes built by atom manipulation ($I = 10$ pA, $V_{\text{DC}} = -100$ mV, image size: 3 nm). The two Fe atoms in the complexes are closest together. **c.** The ESR spectrum measured on Fe site in the right complex marked by red dot in b. ($V_{\text{DC}} = 60$ mV, $V_{\text{rf}} = 12$ mV, $B = 484$ mT, $B_{\text{tip}} = 34$ mT)

The qualitative behavior of the ESR peaks at the point of no detuning $\delta = 0$ can be understood by considering the frequency difference between the second and third ESR resonance, specifically $f_2$ ($f_1$) and $f_3$ ($f_4$) in AFM (FM) case:

$$f_3 - f_2 = \sqrt{\delta^2 + (J_2^{\text{eff}} - D)^2} - (J_2^{\text{eff}} + 2D),$$

$$f_4 - f_1 = \sqrt{\delta^2 + (J_2^{\text{eff}} - D)^2} + (J_2^{\text{eff}} + 2D)$$

Since $|J_2^{\text{eff}}| > D$, we have for $\delta = 0$:

$$\Delta f_2 = f_3 - f_2 = -3D = -3/2 D_0 \quad (J_2^{\text{eff}} > 0, \text{AFM})$$

$$\Delta f_2 = f_4 - f_1 = 3/2 D_0 \ (J_2^{\text{eff}} < 0, \text{FM})$$

Since $D_0 > 0$, $f_3 > f_2$ for the AFM case for $\delta = 0$, which causes the resonances to cross twice as the tip approaches (Fig. S19). In contrast, in the FM case, $f_4 > f_1$, and the two resonances remain separated, as shown in Fig. 4. Thus, crossed or separated transitions at $\delta = 0$ are another indication of AFM and FM coupling, respectively.

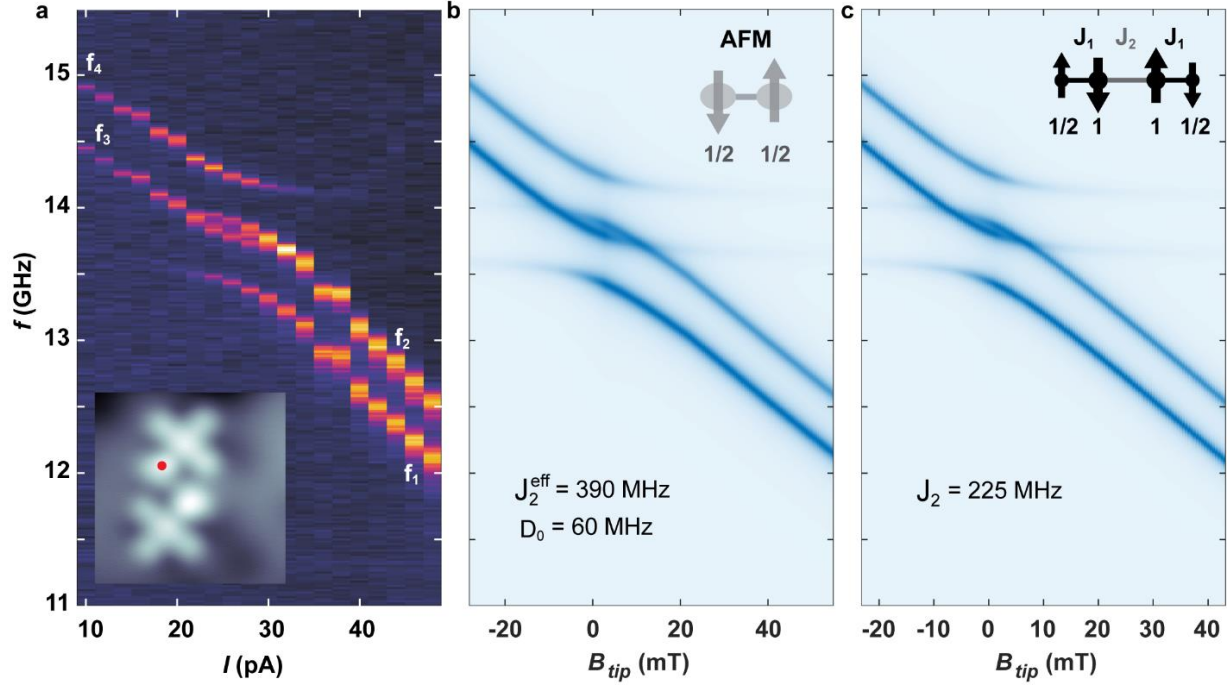

**Supplementary Fig. S19. AFM Coupling in a Dimer of Complexes. a,** ESR measurements at various tip heights, showing an avoided level crossing at $\delta = 0$ ($V_{\text{DC}} = -60$ mV, $V_{\text{rf}} = 12$ mV, $B = 473$ mT). Inset: topographic image ($I = 10$ pA, $V_{\text{DC}} = -100$ mV, image size: 3 nm × 3 nm). **B,** Simulation of the coupled spin system using a two-spin model (inset) of spins (1/2, 1/2), with $J_2^{\text{eff}} = 390$ MHz, $D_0 = 60$ MHz (corresponding to 1.54 nm, roughly (2,5) MgO lattice) and $B = 495$ mT. **c.** Simulation of the coupled spin system using a four-spin model (inset) of spins (1/2, 1, 1, 1/2) with alternating antiferromagnetic coupling strengths $J_1$ and $J_2$. The fitting parameters are: $J_1 = 14.65$ meV, $J_2 = 225$ MHz, $g_{\text{FePc}} = 2$, $g_{\text{Fe}} = 2$, $B = 495$ mT, $T = 1.5\ K$.

We can use the equations of the two spin-model to analyze the data in Fig. 4e and Fig. S19a directly: the magnetic dipolar coupling can be determined by $D_0 = \frac{2}{3}|\Delta f_2|$ (extracted directly at $\delta = 0$) while the inter-complex coupling $J_2^{\text{eff}}$ can be determined by the ESR splitting $\Delta f_1 = J_2^{\text{eff}} + D_0$. For the FM case in Fig. 4, we obtain $D_0 = 131$ MHz, $J_2^{\text{eff}} = -531$ MHz. For the AFM case in Fig. S19, we find $D_0 = 60$ MHz, $J_2^{\text{eff}} = 390$ MHz. The two-spin model is verified through our ESR simulation based on the spin Hamiltonian of two coupled spin ½ (dotted line in Fig. 4e and Fig. S19b). In the simulation, the amplitude of the ESR signal is determined by calculating the matrix element of the transitions, with each state thermally populated according to the Boltzmann distribution of its eigenenergy.

To capture minute details in the position and intensity of the peaks, we employ a simulation of the full four-spin system (Fig. 4f and Fig. S19c), where the coupling between adjacent spins still remains antiferromagnetic and the magnetic dipolar couplings are calculated based on the distances (Fig. S20) between all spin centers. The full four spin model does not show any significant deviations from the two-spin model for both the FM case discussed in the main text and the AFM case discussed in this section. Merely the exchange coupling changes by a factor of around two (Fig. S19), which mostly results from the incorporation of the spin number of the Fe($C_6H_6$) (S=1) into the coupling constant in case of the two-spin model. Thus, we conclude that the complexes can be effectively treated as spin ½ systems and that they can serve as building blocks for larger spin structures.

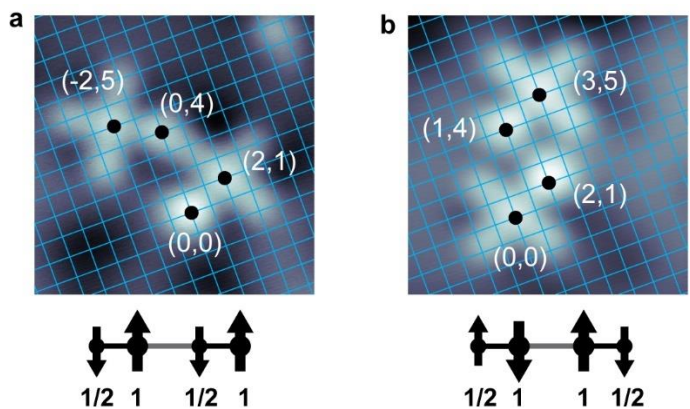

**Supplementary Fig. S20. Lattice fit of two coupled complexes. a.** Topographic image of two ferromagnetically coupled complexes with MgO lattice overlaid as shown in Fig. 4 in the main text. The distances between spins are used for the four-spin model in Fig. 4f. **b.** Topographic image of two antiferromagnetically coupled complexes with MgO lattice overlaid (Fig. S15). The distances between spins are used for the four-spin model in Fig. S15c. Both images: The positions of the Fe and molecule spins are labeled by black dots and are defined relative to the oxygen lattice, which is indicated by a blue grid. Size: 3 nm × 3 nm, setpoint: $V$ = -100 mV, $I$ = 20 pA.